# Energy, Scalability, Data and Security in Massive IoT: Current Landscape and Future Directions

Imane Cheikh[1], Sébastien Roy[1], Essaid Sabir[2], Rachid Aouami[1]

[1]Department of Electrical and Computer Engineering, University of Sherbrooke, Québec, Canada
[2]Department of Science and Technology, TELUQ, University of Quebec, Montreal, H2S 3L4, Canada
Emails: {imane.cheikh, sebastien.roy13, rachid.aouami}@usherbrooke.ca, essaid.sabir@teluq.ca

*Abstract*—The Massive Internet of Things (MIoT) envisions an interconnected ecosystem of billions of devices, fundamentally transforming diverse sectors such as healthcare, smart cities, transportation, agriculture, and energy management. However, the vast scale of MIoT introduces significant challenges, including network scalability, efficient data management, energy conservation, and robust security mechanisms. This paper presents a thorough review of existing and emerging MIoT technologies designed to address these challenges, including Low-Power Wide-Area Networks (LPWAN), 5G/6G capabilities, edge and fog computing architectures, and hybrid access methodologies. We further investigate advanced strategies such as AI-driven resource allocation, federated learning for privacy-preserving analytics, and decentralized security frameworks using blockchain. Additionally, we analyze sustainable practices, emphasizing energy harvesting and integrating green technologies to reduce environmental impact. Through extensive comparative analysis, this study identifies critical innovations and architectural adaptations required to support efficient, resilient, and scalable MIoT deployments. Key insights include the role of network slicing and intelligent resource management for scalability, adaptive protocols for real-time data handling, and lightweight AI models suited to the constraints of MIoT devices. This research ultimately contributes to a deeper understanding of how MIoT systems can evolve to meet the growing demand for seamless, reliable connectivity while prioritizing sustainability, security, and performance across diverse applications. Our findings serve as a roadmap for future advancements, underscoring the potential of MIoT to support a globally interconnected, intelligent infrastructure.

## I. INTRODUCTION

The concept of the Internet of Things (IoT) has rapidly evolved over the last decade, with the number of connected devices expected to surpass 30 billion by 2025 [1]. The emergence of the *Massive Internet of Things* (MIoT) refers to a subset of IoT systems characterized by a vast number of devices generating large amounts of data and requiring highly scalable infrastructures. To support this density of devices and ensure reliable communication, Ultra-Dense IoT Networks (UDNs) have become essential. These networks enable efficient device connectivity and management in environments with extremely high device densities. MIoT deployments, often facilitated by UDNs, are predominantly found in areas such as smart cities, industrial automation, and large-scale environmental monitoring [2], [3].

Scalability in MIoT systems is critical due to the unique demands they impose on communication networks, data processing, energy efficiency, and security. Traditional IoT frameworks are not designed to handle the sheer scale of MIoT in terms of device count, data transmission, and processing requirements [4]. Addressing these scalability challenges is crucial for supporting widespread MIoT applications while maintaining performance, reliability, and security.

One of the key enablers for MIoT scalability is the development of advanced networking protocols such as Low Power Wide Area Networks (LPWANs), including NB-IoT and LoRaWAN, which allow for low-power, long-range communication over large geographical areas [5], [6]. Additionally, emerging computing paradigms such as edge and fog computing are pivotal in offloading computational tasks from centralized cloud platforms to network edges, reducing latency and improving scalability [7].

Another significant challenge in MIoT systems is managing the massive amounts of data generated by connected devices. Efficient data management techniques such as data compression, aggregation, and distributed data processing frameworks (e.g., Hadoop, Spark) are crucial for handling these large datasets [8]. Moreover, energy efficiency remains a key concern, as many MIoT devices are battery-operated and operate in environments with limited access to energy resources. Technologies such as energy harvesting and ultra-low-power communication protocols have emerged as essential solutions for ensuring the sustainability of MIoT deployments [9].

Finally, ensuring the security and privacy of MIoT systems becomes more challenging as the number of devices increases. Traditional centralized security models struggle to scale with the massive device count, prompting the exploration of decentralized solutions like blockchain to secure data transmission and device interactions in MIoT networks [10], [11].

In this paper, we present a comprehensive survey of scalable solutions for MIoT systems, focusing on four critical areas:
- **Network scalability**, where many devices must communicate without overwhelming network infrastructure.
- **Data management**, where vast amounts of data must be processed, stored, and analyzed in real-time.

- **Energy efficiency**, where devices, particularly those in remote or hard-to-reach locations, must operate with limited power resources.
- **Security and privacy**, where ensuring the integrity and safety of vast, distributed networks becomes increasingly complex as the number of devices grows.

We analyze existing approaches and highlight emerging trends that are poised to shape the future of MIoT. Figure 1 presents the main challenges of the massive IoT.

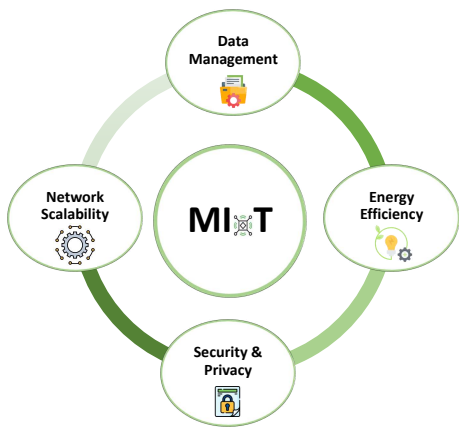

Fig. 1. Massive IoT main challenges.

Furthermore, this paper will also explore *emerging technologies* such as *Artificial Intelligence (AI)* and *Machine Learning (ML)*, which can enhance scalability by optimizing device management, predicting network congestion, and automating decision-making processes. AI-driven solutions can dynamically allocate resources, manage device health, and predict potential system failures before they occur, making large-scale IoT systems more resilient and efficient [12].

As the number of connected devices continues to grow, it is essential to look ahead at the future trends and technological innovations that will drive the next generation of scalable MIoT solutions. From 6G networks [13], which promise even greater capacity and faster speeds, to quantum computing, which could revolutionize data processing, the landscape of MIoT is poised for significant advancements. By understanding the current solutions and the challenges that remain, this paper will contribute to the ongoing conversation about how to scale IoT technologies for a connected future.

In conclusion, this paper aims to provide a detailed exploration of the *technologies, challenges, and solutions* surrounding the scalability of MIoT, highlighting both the current state of the field and the future directions that could define the next wave of innovation. By examining various networking protocols, data management strategies, and emerging technologies, it will offer insights into how MIoT systems can scale effectively while maintaining efficiency, security, and reliability.

*A. Related Surveys*

Throughout the past few years, numerous studies have been made to address the available options for MIoT scalable solutions. Each of these contributions addresses the challenge of network scalability, data delivery, efficiency, security, and privacy, with solutions aimed at these issues.

*1) Network Scalability:* [14] explores the integration of UDNs with other 5G-enabling technologies, such as millimeter waves (mmWave) and massive MIMO. This paper underscores the synergies between these technologies and their potential for addressing scalability and capacity demands in future networks. Similarly, [15] focuses on massive machine-type communications (mMTC) in ultra-dense cellular IoT networks. It investigates the challenges of scalability and resource allocation and proposes machine-learning-assisted solutions to optimize network performance. The survey includes a comprehensive review of techniques like reinforcement learning for managing high device density.

Furthermore, [16] provides a detailed study on mobility management in UDNs. It explores existing solutions and proposes future directions to handle challenges such as handover management and interference mitigation in high-density scenarios. The work [17] categorizes resource allocation techniques for UDNs into different classes and provides a detailed analysis of their applications in 5G and beyond networks. It also identifies open challenges in integrating these techniques with advanced technologies, such as network slicing and AI-based management systems.

An overview of UDNs is presented in [18], discussing their state-of-the-art solutions and potential future directions. It delves into architectural aspects, interference management, and capacity enhancement techniques essential for MIoT scalability. Finally, [19] surveys the evolution of UDNs, emphasizing their role in enabling scalable, efficient, and low-latency communications. The paper provides a critical analysis of existing technologies and suggests novel directions for future research.

A detailed study of the LoRaWAN protocol scalability to cope with mass deployments is also provided in [20]. The survey also addresses the challenges posed by interference in dense environments. It identifies the use of adaptive data rates and interference mitigation techniques as potential solutions to enhance network scalability. Similarly, in [21], the authors reviewed energy-efficient computing for MIoT networks in detail, examining low-power protocols and optimization models that help decrease energy costs in scaled IoT ecosystems.

An additional aspect is offered by [22], which discusses techniques of resource allocation through machine learning guidance so as to achieve improved scalability in ultra-dense cellular IoT networks. This survey identifies key issues, such as resource allocation in high-density areas, which remain a major challenge for MIoT scaling. In this regard, [23] looks into the problems of scalability of data transfer protocols and proposes distribution systems for efficient data transfer within massive IoT systems. These distributed systems are meant to meet the growing demand for scalable data management in MIoT applications.

*2) Data Management:* The survey [30] deals most specifically with IoT big data issues relating to its indexing and attempts to address this problem by proposing distributed databases and real-time data processing approaches for MIoT systems generating enormous quantities of data. This work

TABLE I
COMPARATIVE ANALYSIS OF SCALABLE SOLUTIONS FOR MIOT FROM RECENT SURVEYS.

| Survey | Year of Pub. | Main Topics | Challenges Addressed | Network Scalability | Data Management | Energy Efficiency | Security & Privacy |
|---|---|---|---|---|---|---|---|
| **Our Survey** | 2025 | Integrated framework for scalability, energy efficiency, security | Comprehensive solutions for MIoT challenges | ✓ | ✓ | ✓ | ✓ |
| [21] | 2024 | Energy-efficient computing for MIoT | Low-power protocols, energy optimization | | | ✓ | |
| [20] | 2023 | LoRaWAN scalability, adaptive data rates | Interference mitigation, network scalability | ✓ | | | |
| [24] | 2023 | AoI optimization | Latency, data freshness in MIoT | | ✓ | | |
| [25] | 2023 | Federated learning, deep reinforcement learning | Privacy-preserving data management | | ✓ | | ✓ |
| [26] | 2023 | Blockchain security mechanisms | Lightweight encryption, privacy enhancement | | | | ✓ |
| [27] | 2022 | Resource management in 802.11ah WLAN | Spectrum reuse, network scalability | ✓ | | | |
| [28] | 2021 | Security for MIoT in 6G networks | Privacy-friendly architectures | | | | ✓ |
| [29] | 2021 | Network slicing, AI-driven resource management | Scalability, energy optimization in 6G | ✓ | | ✓ | |
| [17] | 2021 | Resource allocation techniques in UDNs | Scalability, AI-based management systems | ✓ | | ✓ | |
| [30] | 2021 | IoT big data indexing | Real-time processing, data management in MIoT | | ✓ | | |
| [15] | 2020 | mMTC in ultra-dense cellular IoT networks | Scalability, resource allocation with ML | ✓ | | | |
| [14] | 2020 | UDNs with 5G (e.g., mmWave, MIMO) | Scalability, capacity enhancement | ✓ | | | |
| [16] | 2020 | Mobility management in UDNs | Handover management, interference mitigation | ✓ | | | |
| [31] | 2020 | Scalable networking architectures | Load balancing, network flexibility | ✓ | | | |
| [22] | 2019 | Machine learning for resource allocation | Scalability in ultra-dense IoT networks | ✓ | | | |
| [23] | 2019 | Distributed systems, data transfer protocols | Efficient data transfer, scalability | ✓ | ✓ | | |
| [32] | 2017 | Network methodologies for real-time analytics | Massive data generation, IoT scalability | ✓ | ✓ | | |
| [33] | 2018 | Resource allocation in ultra-dense networks | Interference mitigation, network capacity | ✓ | | | |
| [18] | 2016 | UDN solutions overview | Scalability, interference management | ✓ | | | |
| [19] | 2016 | Evolution of UDNs | Scalable, low-latency communication | ✓ | | | |

**Legend**: ✓ indicates the survey's impact in each category.

complements other surveys addressing data management challenges by proposing innovative solutions to handle the ever-increasing data loads in MIoT ecosystems. Also, [24] offers fresh insights on age of information (AoI) within MIoT systems and provides optimization strategies for data freshness and latency problems in the majority of IoT time-critical applications. Another work [25] complements this research by analyzing the potential of deep reinforcement learning for improving federated learning efficiency in MIoT communications, with a focus on privacy-preserving data management.

Another relevant contribution [31] consists in an overview analysis of scalable networking architecture solutions, identifying load balancing and network flexibility as critical issues in MIoT networks. A related effort is [32], which reviews various network methodologies tailored for real-time analytics in MIoT ecosystems. It delves into the challenges posed by massive data generation, transmission, and processing in IoT systems.

*3) Energy Efficiency:* In [21] the authors reviewed energy-efficient computing for MIoT networks in detail, examining low-power protocols and optimization models that help decrease energy costs in scaled IoT ecosystems. Furthermore, [17] provides a comprehensive classification of resource allocation techniques for ultra-dense networks in 5G and beyond, highlighting energy-efficient methodologies for improving network performance. Moreover, [29] presents an extensive study on the role of energy efficiency in the transition towards 6G networks. The work examines optimization strategies aimed at minimizing energy consumption while maintaining reliable and scalable MIoT communications.

*4) Security and Privacy:* In the MIoT, security and privacy remain key areas which have been well studied by a couple of works. [26] presents security mechanisms based on blockchain technologies and decentralized architectures for MIoT systems and emphasizes the need for lightweight encryption for IoT devices in the MIoT context. This line of research is extended by [28], which explores the literature on improving the security of MIoT in 6G networks and providing privacy-friendly architectures based on new technologies.

Our survey builds upon this body of work by providing an integrated framework that addresses the key challenges across four major dimensions: network scalability, data management, energy efficiency, security and privacy. In doing so, it contributes to the ongoing research efforts aimed at enabling scalable, secure, and energy-efficient MIoT deployments. Table 1 presents a comparative analysis of scalable solutions for MIoT from recent surveys.

*B. Research Approach*

In this research, a comprehensive methodology was adopted to explore the scalability challenges and solutions for massive IoT systems. The approach can be divided into the following key phases:

*1) Literature Review and Problem Definition:* An in-depth review of academic publications, industrial reports, and standards from leading organizations such as IEEE, ITU, and ETSI was conducted. The objective was to define the scalability challenges in MIoT and identify existing frameworks and solutions addressing these challenges. The review focused on key areas such as network scalability, data management, energy efficiency, and security/privacy.

*2) Selection Criteria for Scalable Solutions:* The solutions considered were evaluated based on their ability to handle high volumes of IoT devices, flexibility for various use cases, and effectiveness in large-scale deployments. Key evaluation metrics included:

- **Scalability**: The ability to accommodate massive numbers of connected devices.
- **Efficiency**: Optimization of resources such as energy, network bandwidth, and storage.
- **Security and Privacy**: Robustness in protecting data and preventing unauthorized access.

*3) Comparative Analysis:* A comparative analysis was performed by categorizing scalable solutions into networking protocols and architectures, edge and fog computing, and data management techniques. The analysis evaluated factors such as implementation complexity, adaptability, resource efficiency, and trade-offs associated with each solution.

*4) Emerging Trends and Future Directions:* Emerging trends in MIoT, such as 5G networks, blockchain for IoT, and artificial intelligence for adaptive IoT management, were reviewed. Gaps in current solutions were identified, and forward-looking recommendations based on new technological advancements were made.

*C. Motivation*

The unprecedented growth of connected devices in the MIoT domain presents significant challenges to existing network infrastructures. Forecasts suggest the number of IoT devices will surpass 75 billion by 2025, necessitating highly scalable, efficient, and secure frameworks. These challenges include network scalability, efficient data management, energy conservation, and robust security measures, forming the foundation of this study.

**Network Scalability** has become a critical concern as conventional networking architectures are unable to accommodate the enormous device density expected in IoT ecosystems. Current research highlights advanced strategies to address these issues. For example, Ramachandran et al. (2019) propose hybrid communication architectures that combine LPWAN with 5G technologies to enhance scalability and meet the growing demand for connectivity in dense IoT networks [34]. Furthermore, Khan et al. (2020) investigate network slicing techniques to allocate resources dynamically, ensuring optimal performance in diverse IoT applications [35].

**Data Management** is another pressing issue, given the vast amounts of data generated by IoT devices. Existing storage and processing solutions often fall short of the requirements for real-time analytics and decision-making. Modupe et al. (2024) emphasize the potential of edge computing frameworks in decentralizing data processing, reducing latency, and enabling rapid responses in time-sensitive IoT applications [36]. In addition, advancements in AI-driven analytics, as discussed by Zong et al. (2024), have further enhanced the ability to derive insights from large datasets without overburdening centralized servers [37].

**Energy Efficiency** remains a crucial aspect, particularly as many IoT devices operate with limited power supplies. Research by Tupe et al. (2022) focuses on energy-aware communication protocols that optimize power consumption without compromising data integrity, thereby prolonging device lifespans [38]. Similarly, studies on energy harvesting technologies show promise in making MIoT networks more sustainable by utilizing ambient energy sources [39].

Lastly, **Security and Privacy** challenges have become increasingly pronounced with the rise in connected devices. The growing attack surface demands multi-layered security frameworks. Cui et al. (2019) propose novel blockchain-based architectures to secure IoT environments while ensuring transparency and data immutability [40]. Meanwhile, AI-enabled intrusion detection systems are gaining traction for their ability to detect and mitigate sophisticated threats in real time [41].

In conclusion, this research aims to address these challenges by proposing innovative solutions that integrate scalability, efficient data handling, energy optimization, and robust security into a unified MIoT framework. By leveraging cutting-edge technologies and recent advancements, this study seeks to contribute to the development of sustainable and reliable IoT systems.

### D. Contributions

This paper provides a comprehensive analysis of the scalability challenges and solutions in the context of MIoT, contributing to both academic understanding and practical implementations along the following axes:

- **Holistic Review of Scalable IoT Solutions**: We offer a broad review of existing scalable solutions in networking protocols, data management techniques, energy-efficient designs, and security frameworks. This review consolidates knowledge across multiple domains, providing a unified perspective on scalability in MIoT.
- **Comparative Analysis of Solutions**: A detailed comparative analysis is presented, highlighting the trade-offs between different approaches. This analysis is structured around key metrics such as scalability, efficiency, security, and cost-effectiveness. Our findings serve as a reference for choosing appropriate solutions based on specific MIoT application requirements.
- **Identification of Gaps and Emerging Trends**: We identify significant gaps in current solutions, such as limitations in security frameworks and energy inefficiency in large-scale deployments. In addition, we highlight emerging trends like edge and fog computing, 5G, and AI-powered IoT management, offering insights into how these technologies can address existing limitations.
- **Framework for Future Research**: Based on our review and analysis, we propose a framework for future research on scalable IoT systems, focusing on the integration of advanced technologies such as blockchain for secure IoT transactions, and hybrid cloud-edge architectures for efficient data processing. This framework provides guidelines for the development of next-generation IoT solutions.

Overall, this paper provides a roadmap for overcoming scalability challenges in MIoT, offering both theoretical insights and practical guidelines for researchers and practitioners.

### E. Organization

The remainder of this paper is organized as follows:

In **Section II**, we present various *Scalable MIoT Use Cases*, highlighting the practical applications of massive IoT across industries such as smart cities, healthcare and agriculture. This section outlines the benefits of scalability in these areas and their specific challenges.

**Section III** addresses the *Scalability Challenges in MIoT*, focusing on the key challenges of network scalability, data management, energy efficiency, and security & privacy. Additionally, a *Comparative Analysis of MIoT* is included to provide an in-depth evaluation of existing solutions against these challenges.

**Section IV** presents *Advanced Techniques for Scalable MIoT Networks*, where various advanced techniques are examined, including grant-free access, cooperative communication, and NOMA (Non-Orthogonal Multiple Access). This section delves into how these techniques contribute to enhancing network scalability, energy efficiency, and secure communication in large-scale MIoT deployments.

**Section V** explores Advanced Tools for Scalable MIoT Networks, where cutting-edge methodologies and technologies are analyzed for optimizing MIoT deployments. This section covers the role of artificial intelligence (AI) and machine learning (ML) in optimizing resource allocation and network adaptability.

**Section VI** explores the *Emerging Trends and Future Solutions* in MIoT. This section discusses the latest advancements and trends, such as artificial intelligence, edge computing, blockchain technology, and the anticipated impact of 6G networks on MIoT scalability.

Finally, **Section VII** presents the *Conclusions* of this work, summarizing the key findings and discussing the future direction of scalable MIoT solutions. Gaps in current solutions are highlighted and relevant avenues for further research and innovation are identified. Figure 2 provides a general overview of the paper structure, while acronym definitions are presented in Table 2.

## II. SCALABLE MIoT USE-CASES

The concept of Massive IoT encompasses diverse applications across various sectors, each presenting unique scalability requirements and challenges. The following sections outline specific use cases, illustrating how scalable solutions in MIoT can drive efficiency and innovation in these industries.

### A. Smart cities

Smart cities use MIoT solutions to enhance urban infrastructure, sustainability, and quality of life. MIoT applications in these cities range from adaptive traffic management, smart lighting, waste management, air quality monitoring, and emergency response systems, all relying on real-time data

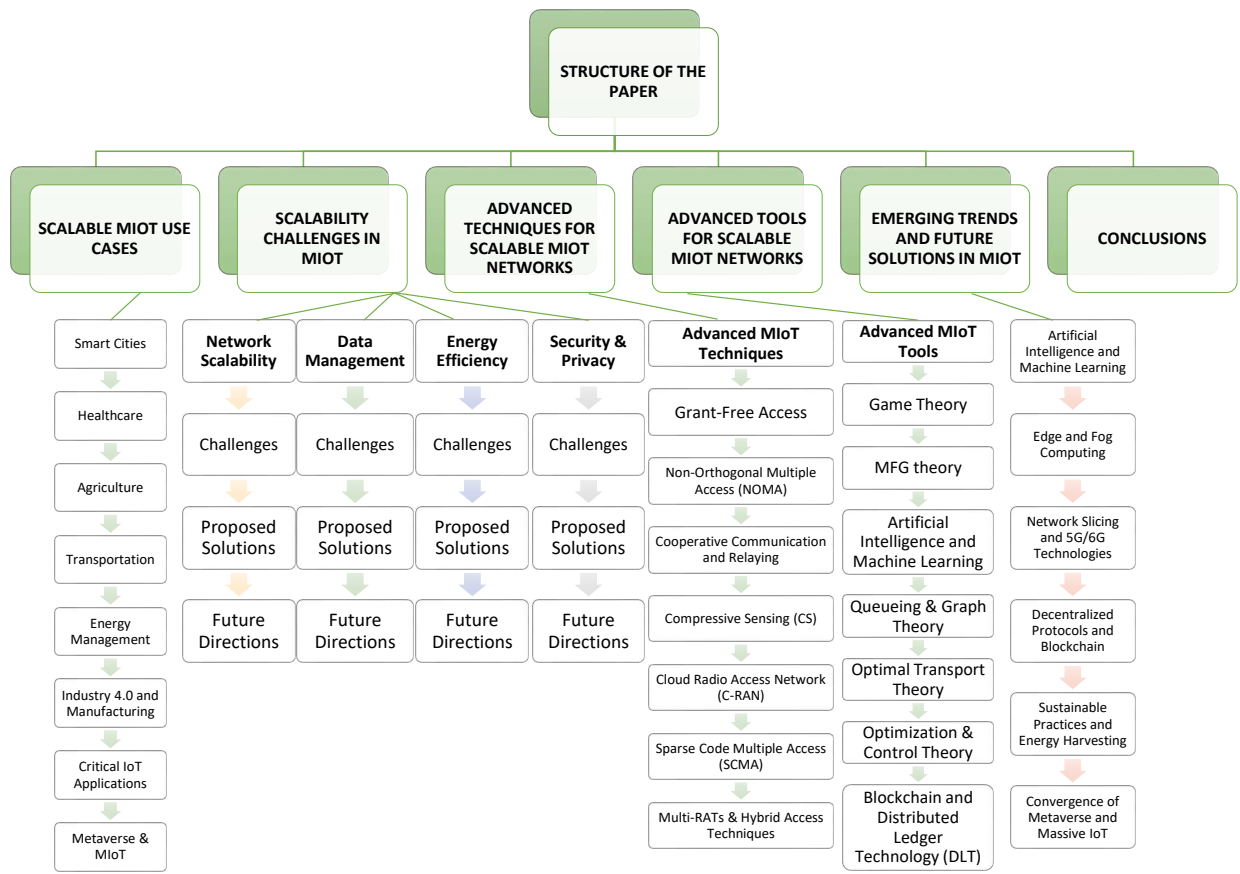

Fig. 2. Structure of the paper.

from interconnected IoT devices to deliver timely insights and optimize operations [42]–[44].

A prominent example is adaptive traffic control, where sensors at intersections monitor vehicle flow and congestion patterns, and traffic signals adjust in real-time to reduce bottlenecks and improve traffic flow efficiency. This approach can cut down commute times by up to 30% in densely populated areas by processing data at the network edge, thus minimizing latency and ensuring faster response times [45], [46]. Edge computing frameworks integrated with such systems play a critical role in ensuring low-latency data processing and reducing reliance on centralized servers, which is essential for handling the vast amounts of data generated by urban IoT networks [47], [48].

Smart lighting systems also contribute significantly to energy savings in urban areas by adjusting light levels based on occupancy and environmental data. Studies indicate that implementing such systems can reduce energy consumption by nearly 50%, enhancing sustainability and reducing operational costs [49], [50]. Some advanced systems leverage AI-powered sensors and adaptive lighting algorithms to maintain optimal illumination only when necessary, creating a more sustainable urban infrastructure. Furthermore, waste management systems in smart cities now use IoT-enabled sensors on waste bins to track fill levels and optimize collection routes, which reduces fuel consumption and operational costs for waste management services [51], [52]. Such MIoT systems also allow cities to monitor air quality, noise pollution, and water quality, integrating data streams to form a comprehensive environmental monitoring network. This data not only improves municipal services but also empowers citizens to make informed decisions about health and well-being [53], [54].

Overall, smart cities benefit from MIoT solutions by achieving greater scalability, enhanced energy efficiency, and improved data management. However, challenges remain, particularly in managing the vast data generated and ensuring security and privacy across all interconnected devices.

### B. Healthcare

The integration of MIoT applications in healthcare is revolutionizing patient monitoring, telemedicine, and health data management, providing solutions that are both scalable and patient-centered [55], [56]. Wearable devices, such as heart rate monitors and glucose sensors, continuously monitor patient vitals, transmitting data to healthcare providers for timely assessment and intervention. These devices allow early detection of health issues and minimize the need for in-person consultations, which is particularly beneficial for patients with chronic illnesses [57].

Managing the large volumes of health data generated by these devices poses challenges in data scalability, privacy, and network load. To address these, federated learning has been widely adopted to process data locally on devices, thus preserving patient privacy and reducing network dependency

TABLE II
LIST OF ABBREVIATIONS

| Abbreviation | Definition |
|---|---|
| 5G | Fifth Generation |
| 6G | Sixth Generation |
| AF | Amplify-and-Forward Scheme |
| AI | Artificial Intelligence |
| AIS | Artificial Immune Systems |
| AoI | Age of Information |
| AR | Augmented Reality |
| BBU | Baseband Unit |
| BFT | Byzantine Fault Tolerance |
| CRAN | Cloud-Radio Access Network |
| CRN | Cognitive Radio Network |
| CS | Compressive Sensing |
| D2D | Device-to-Device Communication |
| DAG | Directed Acyclic Graph |
| DERs | Distributed Energy Resources |
| DF | Decode-and-Forward Scheme |
| DL | Deep Learning |
| DLT | Distributed Ledger Technology |
| DPoS | Delegated Proof of Stake |
| DSA | Dynamic Spectrum Access |
| eMBB | Enhanced Mobile Broadband |
| FL | Federated Learning |
| IDS | Intrusion Detection Systems |
| IIoT | Industrial Internet of Things |
| IoT | Internet of Things |
| LPWAN | Low-Power Wide Area Network |
| MAC | Medium Access Control |
| MC-NOMA | Multi-Carrier Non-Orthogonal Multiple Access |
| MEC | Multi-access Edge Computing |
| MFG | Mean Field Game |
| MFGT | Mean Field Game Theory |
| MIoT | Massive Internet of Things |
| ML | Machine Learning |
| MLP | Machine Learning Processing |
| mMTC | Massive Machine Type Communication |
| mmWave | Millimeter-Wave |
| Multi-RAT | Multi-Radio Access Technology |
| NFV | Network Function Virtualization |
| NOMA | Non-Orthogonal Multiple Access |
| NOMA-OFDMA | Non-Orthogonal Multiple Access with Orthogonal Frequency Division Multiple Access |
| OT | Optimal Transport |
| P2P | Peer-to-Peer Communication |
| PoS | Proof of Stake |
| PoW | Proof of Work |
| QoS | Quality of Service |
| RAN | Radio Access Network |
| RL | Reinforcement Learning |
| RRH | Remote Radio Head |
| SCMA | Sparse Code Multiple Access |
| SDN | Software-Defined Networking |
| SIC | Successive Interference Cancellation |
| THz | Terahertz |
| TTI | Transmission Time Interval |
| UAV | Unmanned Aerial Vehicle |
| URLLC | Ultra-Reliable Low-Latency Communication |
| VR | Virtual Reality |
| WLAN | Wireless Local Area Network |
| XR | Extended Reality |

[58], [59]. This approach also enables continuous updates to AI models, improving healthcare delivery through adaptive learning without compromising sensitive patient information [60]. AI further enhances MIoT healthcare applications by analyzing massive datasets to identify patterns that may assist in proactive health management and predictive diagnosis [61], [62]. For example, AI algorithms can help identify early signs of diseases such as diabetes or cardiovascular conditions based on historical data, providing actionable insights for healthcare providers [63]. Additionally, advances in edge computing have facilitated faster data processing on wearable devices, which is essential for real-time health applications requiring low latency and reliability [64]. These technologies collectively foster a healthcare environment where MIoT systems can manage high device densities, ensure data security, and provide scalable solutions suitable for personalized and preventive healthcare.

*C. Precision Agriculture*

Precision agriculture leverages MIoT solutions to enhance productivity by using data from IoT sensors and remote sensing devices. These sensors, deployed across large agricultural fields, collect critical data on soil moisture, temperature, humidity, and nutrient levels, enabling farmers to make precise, data-driven decisions regarding irrigation, fertilization, and pest control [65]–[67]. Additionally, data from these sensors combined with weather forecasts and crop growth models provide valuable insights for optimizing resource allocation and minimizing waste, thereby supporting sustainable farming practices. As sensor networks expand in scale, maintaining connectivity becomes increasingly challenging, especially in remote areas. LPWANs like LoRaWAN and NB-IoT, alongside energy-efficient sensors, are crucial for scalability in these networks, enabling data transmission over vast distances with minimal energy consumption [68], [69]. In areas with limited connectivity, edge computing frameworks allow data to be processed locally, reducing the need for continuous data transmission and ensuring timely decisions even in remote settings [70].

Emerging AI-driven analytics are also proving beneficial in precision agriculture. By analyzing large-scale data from sensors, AI models can predict crop yields, optimize water usage, and detect pest infestations early. These insights lead to improved productivity and environmental sustainability [71], [72]. For example, integrating machine learning models with MIoT systems helps optimize fertilizer application rates based on real-time soil data, reducing over-fertilization and its environmental impact [73]. This combination of scalable MIoT infrastructure, energy-efficient networks, and AI analytics represents a transformative shift in agricultural practices, enabling farmers to enhance yields while conserving resources and promoting ecological sustainability.

*D. Intelligent Transportation*

MIoT applications in transportation are essential to the advancement of autonomous vehicles, real-time logistics, and fleet management. Autonomous vehicles (AVs) utilize MIoT infrastructure to communicate with nearby vehicles, road infrastructure, and environmental sensors, which is crucial for navigating complex urban environments and improving safety [74]–[76]. These applications generate substantial volumes of data, which need to be processed with high speed and minimal latency to ensure timely and reliable operation. As the number of connected vehicles grows, the scalability of MIoT solutions becomes critical to maintaining efficient and safe transport systems. To meet these high data and communication demands, MIoT in transportation relies on 5G technology and multi-access edge computing (MEC), which together support low-latency, high-throughput data processing at the network edge. MEC enables data to be processed closer to the AVs, reducing the need for continuous back-and-forth communication with centralized servers [77], [78]. Research shows that integrating 5G with MEC can significantly improve the responsiveness of AVs in high-density traffic, enhancing safety and scalability in urban settings [79], [80].

Furthermore, MIoT facilitates real-time logistics and fleet management. Logistics companies are using sensor-equipped vehicles and AI-driven analytics to track location, monitor vehicle health, and optimize delivery routes, leading to cost savings and improved delivery times [81], [82]. Predictive maintenance, enabled by MIoT sensors, allows fleet operators to monitor vehicle conditions in real-time, reducing downtime and improving operational efficiency [83]. As the number of connected vehicles continues to increase, scalable MIoT frameworks are essential for ensuring these systems function seamlessly across various urban and rural environments.

*E. Energy Management*

MIoT applications in energy management, particularly within smart grids, are revolutionizing how energy distribution and reliability are maintained by integrating real-time data from IoT-enabled devices, such as smart meters, substations, and energy management systems [84], [85]. These systems collect and analyze data to enable utilities to forecast energy demand, balance loads, reduce outages, and optimize the integration of renewable energy sources, thereby advancing sustainability goals. By leveraging MIoT, energy providers can enhance grid flexibility and efficiency, which is essential for meeting growing energy demands in urban and rural areas.

Scalability in MIoT-driven energy management is further enhanced by predictive algorithms that process historical consumption data, enabling accurate forecasting and effective demand-response actions [86], [87]. ML techniques contribute to the robustness of smart grids by supporting predictive maintenance, which can preempt equipment failures and reduce downtime [88]. Moreover, integrating renewable sources, such as solar and wind, into these grids requires adaptive MIoT architectures that can manage intermittent supply and fluctuating demand. Distributed energy resources (DERs) are increasingly used alongside MIoT-enabled devices for efficient energy storage and real-time distribution adjustments [89].

Research indicates that edge computing frameworks in MIoT can further improve the scalability and resilience of smart grids by reducing latency in data processing, thereby enabling faster and more localized decision-making [90]. Additionally, blockchain technology is being explored to enhance data security within energy management systems, providing tamper-resistant records of energy transactions and ensuring trust among distributed energy producers and consumers [91].

*F. Industry 4.0 and Manufacturing*

The adoption of MIoT in manufacturing is foundational to Industry 4.0, enabling the creation of smart factories that leverage interconnected devices for monitoring production lines, managing inventories, and optimizing processes [92], [93]. MIoT devices continuously gather and analyze data on machine performance, facilitating predictive maintenance that reduces unexpected downtime, minimizes repair costs, and enhances productivity [94]. Predictive maintenance is critical in manufacturing environments, where minor interruptions in the production process can lead to significant costs and delays.

Scalability in Industry 4.0 is paramount, as these environments often require supporting thousands of sensors, actuators, and controllers within a single factory [95]. High data volume and the need for low latency make it essential to deploy scalable solutions such as LPWANs, which provide long-range communication for connected devices, and hybrid cloud-edge architectures that reduce latency by processing data closer to the source [96], [97]. These architectures not only allow for efficient data handling, but also improve data security and resilience by distributing processing across multiple layers.

MIoT's role in manufacturing extends to real-time quality control and inventory management. Sensors monitor environmental factors like temperature and humidity, ensuring products meet quality standards throughout production [98]. Additionally, automated inventory systems integrated with MIoT provide up-to-the-minute stock information, enhancing the ability to manage supply chains and reduce waste [99].

Future advancements in MIoT for manufacturing must focus on enhancing scalability and adaptability to support evolving Industry 4.0 requirements. These innovations include integrating artificial intelligence for adaptive process management and implementing advanced cyber-security measures to protect sensitive industrial data [100], [101]. As MIoT continues to develop, its role in transforming manufacturing will rely on adaptable architectures that can meet the dynamic demands of data-driven, interconnected industrial environments.

*G. Critical IoT Applications*

The advent of 5G technology has introduced a new communication paradigm known as Ultra-Reliable Low-Latency Communication (URLLC), which is specifically designed to meet the stringent requirements of critical IoT applications. Critical IoT use cases, such as industrial automation, autonomous vehicles, remote surgery, and smart grid management, demand ultra-high reliability, extremely low latency, and consistent throughput. These requirements are essential to ensure the safe and efficient operation of systems where even minor delays or failures can have catastrophic consequences [102].

The MIoT, consisting of tens of billions of connected devices, objects, and machines, relies on ubiquitous connectivity. Among them, certain IoT devices require URLLC [103].

URLLC achieves reliability levels of up to 99.999% through techniques like diversity, redundancy, and advanced error correction [104], while ensuring end-to-end latency as low as 1 millisecond via edge computing, short transmission time intervals (TTIs), and preemptive scheduling [105]. Additionally, constant throughput is maintained through dynamic resource allocation, QoS guarantees, and technologies like beamforming and massive MIMO. Despite its potential, challenges such as network synchronization, energy efficiency, and scalability must be addressed to fully realize URLLC's capabilities in supporting the next generation of critical IoT systems [106].

### H. Metaverse and Massive IoT

The convergence of the Metaverse and MIoT is poised to revolutionize various industries by creating immersive, interactive, and data-rich environments. MIoT, characterized by the extensive deployment of interconnected devices, serves as a foundational layer for the Metaverse, enabling real-time data integration from the physical world into virtual spaces.

In healthcare, this integration facilitates advanced telemedicine and remote monitoring. IoT devices collect patient data, which can be visualized in the Metaverse, allowing healthcare professionals to assess and interact with patient information in a 3D virtual environment. This approach enhances diagnostic accuracy and patient engagement [107].

The retail sector also benefits from the fusion of MIoT and the Metaverse. IoT sensors track inventory and consumer behavior, while virtual platforms offer immersive shopping experiences. Customers can virtually try on products or explore digital showrooms, bridging the gap between online and physical retail [108].

In industrial applications, the concept of the industrial metaverse emerges, where IoT devices monitor machinery and processes, and digital twins replicate these systems in virtual environments. This setup allows for real-time monitoring, predictive maintenance, and optimization of operations, leading to increased efficiency and reduced downtime [109].

Moreover, the integration of IoT with the metaverse enhances smart city initiatives. IoT devices collect data on urban infrastructure, which can be visualized in the metaverse to simulate and manage city operations, improving resource allocation and urban planning [107].

In summary, the synergy between the metaverse and MIoT offers transformative potential across various sectors, enabling more interactive, efficient, and data-driven applications.

## III. SCALABILITY CHALLENGES IN MIoT

MIoT introduces unique scalability challenges across several dimensions—ranging from network architecture, data management, and energy efficiency to security and privacy concerns. The high density amplifies these challenges, in heterogeneous environments in which MIoT systems operate. Addressing these issues requires innovative approaches that balance performance, energy consumption, and data security while supporting massive connectivity and real-time processing.

In the following sections, we explore the key scalability challenges and potential solutions in MIoT environments.

### A. Ultra-Dense IoT Networks

Network scalability is a fundamental aspect of MIoT systems, as it determines the ability of the network to support an increasing number of devices while maintaining performance, reliability, and efficiency. The rapid growth in IoT devices necessitates the development of scalable network architectures that can handle high device density, ensure low latency, manage resources efficiently, and handle interference effectively.

*1) Challenges of Network Scalability:* Scaling IoT networks presents several challenges, especially in high-density environments:

- **Device Density and Network Congestion**: As the number of connected devices increases, network congestion becomes a critical issue. Without proper management, congestion can lead to packet loss, increased latency, and reduced overall network performance. Research suggests that novel congestion control techniques are required to ensure smooth network operations in dense IoT environments [118], [119].
- **Latency Requirements**: Many IoT applications, particularly those requiring real-time processing such as autonomous vehicles and industrial automation, are highly sensitive to latency. High device densities can lead to increased traffic, resulting in longer delays. Effective techniques for managing latency, including optimized routing protocols and edge computing, are essential for meeting the stringent requirements of these applications [120], [121].
- **Resource Allocation**: Dynamic and efficient resource allocation is crucial for ensuring scalability in IoT networks. With the growing number of devices, allocating resources such as bandwidth, processing power, and memory becomes increasingly complex. Recent research highlights the importance of AI-driven solutions for dynamically managing resources to avoid network bottlenecks and enhance scalability [122], [123].
- **Interference Management**: In dense MIoT deployments, interference from overlapping wireless signals can degrade communication quality. Effective interference management strategies, such as adaptive power control and interference-aware protocols, are necessary to maintain reliable communication as the network scales [124], [125].
- **Protocol Scalability**: Traditional networking protocols, such as those used in legacy systems, were not designed to support the scale of MIoT. New or adapted protocols that can handle a large number of devices while minimizing signaling overhead and energy consumption are being researched and developed [126], [127].

*2) Contemporary solutions:* To address these challenges, several technological solutions have been proposed:

TABLE III
COMPARISON OF NETWORK SCALABILITY APPROACHES IN MIoT NETWORKS.

| Technology/Approach | Key Features | Advantages | Challenges/Limitations | References |
|---|---|---|---|---|
| Cognitive Radio Networks (CRN) | Dynamic spectrum allocation based on real-time traffic patterns | Maximizes spectrum utilization; reduces congestion in dense networks | Requires sophisticated sensing and decision-making mechanisms | [110] |
| Millimeter-Wave (mmWave) | Utilizes higher frequency bands (30-300 GHz) | Vast bandwidth availability; supports high data rates | High path loss; limited range and penetration | [111] |
| NB-IoT & LTE-M (Cellular IoT) | Low-power, wide-area coverage; supports large device densities | Supports millions of devices per km²; low power consumption | Limited bandwidth; congestion in ultra-dense areas | [112] |
| LoRaWAN (LPWAN) | Long-range, low-power communication | Excellent for low-power, long-range IoT applications | Scalability issues due to limited bandwidth and collision risks | [113] |
| Software-Defined Networking (SDN) | Centralized network management and resource allocation | Flexible, scalable, and supports dynamic traffic management | Security concerns; overhead in managing large-scale IoT | [114] |
| Routing Protocols (RPL) | Energy-efficient routing protocol for low-power IoT devices | Reduces routing overhead; supports thousands of nodes | May struggle with high mobility scenarios | [115] |
| Machine Learning-Assisted Scalability | Predicts traffic patterns; optimizes network resources dynamically | Improves performance and resource allocation efficiency | Requires extensive training data and computational resources | [116] |
| Edge/Fog Computing | Distributes computing resources close to data sources | Reduces latency; minimizes load on central servers | Complexity in management and deployment | [117] |

- **Cognitive Radio Networks (CRN)**: CRN enables dynamic spectrum access, allowing IoT devices to utilize unused frequency bands opportunistically. This approach significantly reduces congestion and improves the network's support of large-scale deployments. Cognitive techniques have shown promising results in optimizing network performance in dense environments [128], [129].
- **Millimeter-Wave (mmWave) Communications**: mmWave technology offers large bandwidths, enabling high data rates and accommodating many simultaneous connections. Although the propagation characteristics of mmWave signals lead to limited range and require advanced beamforming techniques, they offer a significant potential for scaling IoT networks in urban areas [130], [131].
- **Low-Power Wide-Area Networks (LPWANs)**: LPWAN technologies such as LoRaWAN and NB-IoT are designed to provide long-range, low-power communication, making them ideal for IoT networks with many devices. While these technologies excel in energy efficiency, addressing scalability in terms of data throughput remains a challenge [132], [133].
- **Software-Defined Networking (SDN)**: SDN allows for the decoupling of the control plane from the data plane, providing more flexibility and scalability in managing network resources. SDN-based approaches enable centralized control, allowing the network to dynamically adapt to varying traffic conditions and device densities [134], [135].
- **Edge and Fog Computing**: Edge and fog computing architectures push data processing closer to IoT devices, reducing latency and bandwidth consumption. By enabling real-time data processing at the edge of the network, these computing paradigms are essential for scaling MIoT systems [136], [137].

Table 3 presents a comparative analysis of network scalability approaches in MIoT networks.

*3) Future Research Directions:* Future research on network scalability should focus on integrating these technologies to develop hybrid solutions. For example, combining cognitive radio techniques with SDN may improve both spectrum management and resource allocation. Additionally, incorporating machine learning techniques for predictive traffic and resource management could further enhance scalability in MIoT networks [138], [139]. Addressing these open challenges will pave the way for more resilient and efficient MIoT systems.

Network scalability is a fundamental challenge for the future of MIoT, as the number of connected devices continues to grow exponentially. Addressing issues related to device density, bandwidth limitations, and varying device requirements is critical for the continued expansion of IoT. While current solutions such as LPWAN, 5G, and mesh networks provide the backbone for scaling MIoT deployments, ongoing innovations, including 6G and satellite IoT networks, will be vital in ensuring the long-term scalability and success of the massive IoT landscape.

*B. Data Management*

MIoT systems involve billions of connected devices that generate vast amounts of data. Managing this data, in terms of storage, processing, analysis, and security, presents several challenges. Effective data management is essential for ensuring the scalability, reliability, and security of MIoT environments. This section highlights the primary challenges and proposed solutions for efficient data management in MIoT systems.

*1) Challenges in Data Management:* The large-scale nature of MIoT systems introduces various challenges:

**1. Data Volume:** MIoT environments generate an immense amount of data. For example, smart cities, with connected devices monitoring everything from traffic to air quality, produce terabytes of data per day. Storing and processing this data in traditional centralized cloud infrastructures leads

TABLE IV
COMPARISON OF PROPOSED SOLUTIONS FOR DATA MANAGEMENT IN MASSIVE IOT.

| Proposed Solution | Scalability | Efficiency | Security | Advantages | Challenges | Example of Technique | References |
|---|---|---|---|---|---|---|---|
| Edge computing architectures | *** | *** | ** | Reduces latency and bandwidth usage | Latency and bandwidth constraints | Local processing for real-time analytics | [117], [140] |
| Data aggregation techniques | ** | *** | ** | Lowers data transfer volume, reduces processing load | Data loss during aggregation | Hierarchical clustering for data compression | [141], [142] |
| Fog computing approaches | *** | ** | *** | Distributes computing resources, reduces cloud dependency | Complexity in deployment | Multi-tier processing to offload tasks from cloud | [143], [144] |
| Cloud-based data management | **** | *** | * | Provides high storage and processing capabilities | Data privacy concerns | Centralized data storage for extensive analytics | [145], [146] |
| Machine learning for data analytics | ** | *** | *** | Enhances predictive accuracy and pattern recognition | Model training, potential biases | Anomaly detection for device performance | [147], [148] |

**Legend:** * = Low, ** = Medium, *** = High, **** = Very High

to inefficiencies due to bandwidth constraints and latency [149]. Centralized systems become bottlenecks as the data load increases, demanding distributed data storage techniques [150].

**2. Data Variety:** The variety of data generated in MIoT systems is another key challenge. Devices produce structured, semi-structured, and unstructured data, including sensor readings, video streams, audio signals, and logs. Integrating and managing such heterogeneous data within a single framework is difficult. Furthermore, different types of data require different methods of processing and storage, increasing system complexity. Big data technologies, such as Hadoop and Spark, are widely adopted to handle this diversity, providing flexible and scalable platforms [151].

**3. Data Velocity:** In addition to high volume, MIoT applications often require real-time or near real-time data processing. For instance, healthcare applications relying on IoT data for remote patient monitoring must process and analyze data continuously to enable immediate responses. This need for low-latency processing in the face of high data velocity is another core challenge in MIoT [152]. Traditional batch processing systems are unsuitable for these use cases, leading to the rise of stream processing systems and edge computing solutions [153].

**4. Data Integrity and Security:** The vast number of devices connected to MIoT systems increases the potential attack surface for cyber threats. Ensuring the security of data throughout its lifecycle—from collection to transmission and storage—is a major concern. In addition, ensuring the integrity of data, particularly when transmitted over unreliable networks, is essential for maintaining trust in IoT applications. Technologies like blockchain and encryption algorithms are crucial in addressing these security concerns [154], [155].

**5. Energy Efficiency:** Many IoT devices operate with limited computational power and energy resources. Efficiently managing data without draining device batteries is essential. Protocols that minimize the energy consumed during data transmission and processing are required to ensure the long-term viability of MIoT networks [156].

*2) Contemporary solutions:* Given these challenges, researchers have proposed several solutions:

**1. Edge and Fog Computing:** Edge computing involves processing data closer to the source (i.e., at the edge of the network), reducing latency and bandwidth requirements. This approach allows for data to be processed in real-time, alleviating the load on centralized cloud systems and reducing the distance that data must travel [157]. Fog computing extends the edge concept by enabling intermediate nodes between the cloud and edge devices, allowing for more distributed and scalable processing [158].

**2. Federated Learning:** In traditional centralized machine learning, data from IoT devices must be transferred to a central server for model training, which poses privacy and bandwidth issues. Federated learning mitigates this problem by training machine learning models locally on devices and sharing only the model updates, not the raw data. This approach enhances data privacy, reduces bandwidth consumption, and improves scalability [159]. Federated learning is particularly well-suited for MIoT environments, where privacy and energy efficiency are top concerns.

**3. Blockchain for Data Security:** Blockchain technology has emerged as a promising solution for enhancing security in distributed MIoT systems. Its decentralized and immutable ledger ensures that data is securely stored and can be audited without the risk of tampering [160]. Blockchain also supports secure data exchange between IoT devices, enhancing trust in peer-to-peer networks and reducing reliance on central authorities [161].

**4. Data Compression and Aggregation:** Data compression techniques can reduce the size of the data transmitted by IoT devices, addressing bandwidth limitations. Lossless compression methods maintain data integrity, while lossy techniques can be employed for applications where some data loss is acceptable. In addition, data aggregation at the edge reduces redundancy and ensures that only relevant data is transmitted, optimizing both energy consumption and network usage [162].

**5. Real-Time Data Processing Frameworks:** To handle high-velocity data, real-time processing frameworks, such as Apache Kafka and Spark Streaming, have been adopted in MIoT. These systems provide scalable solutions for processing continuous data streams and generating actionable insights on-the-fly, which is crucial for applications like smart cities, healthcare, and industrial automation [163].

*3) Future Directions:* Future research in MIoT data management is expected to focus on further optimizing scalability, security, and energy efficiency. Integrating AI and machine learning techniques into data management frameworks will play a pivotal role in handling complex data streams, optimizing resource allocation, and predicting system behavior. Privacy-preserving techniques, such as homomorphic encryption and differential privacy, will be crucial in ensuring secure data processing in federated learning environments [164], [165]. Additionally, the emergence of 6G networks will provide the ultra-low latency and high bandwidth needed to support the next generation of MIoT systems [166].

Table 4 summarizes several proposed solutions for data management in MIoT, focusing on their scalability, efficiency, security features, and the challenges they address.

### C. Energy Efficiency

Energy efficiency is a critical aspect of the MIoT, as it directly influences the longevity, reliability, and scalability of IoT deployments. The large number of low-power devices operating in MIoT networks necessitates the development of energy-efficient communication protocols, data processing techniques, and resource management solutions. This section provides an in-depth review of the energy challenges faced by MIoT systems and the corresponding solutions and future research directions.

*1) Challenges in Energy Efficiency:* The scale and complexity of MIoT present several challenges to achieving energy efficiency:

- **Limited Power Resources:** Most IoT devices rely on batteries with limited lifespans, and the deployment of these devices in remote or inaccessible areas makes battery replacement costly and impractical [167].
- **Massive Data Generation:** The high volume of data generated by MIoT devices increases energy consumption during data transmission and processing, especially when cloud-based storage and processing are involved [168].
- **Network Scalability:** As the number of connected devices grows, managing the energy consumption of each device and ensuring efficient operation becomes a major challenge, particularly in heterogeneous networks with varying device capabilities [169], [170].
- **Interference and Congestion:** IoT devices operating in dense environments often experience interference and network congestion, leading to increased energy consumption due to repeated transmission attempts [171].
- **Sustainability:** Achieving energy efficiency at scale is crucial for the long-term sustainability of MIoT ecosystems, as large-scale deployments are expected to grow exponentially with the advent of 5G and 6G technologies [172], [173].

*2) Contemporary solutions:* Various solutions have been proposed to address these challenges, focusing on minimizing energy consumption while maintaining performance and scalability:

- **Energy Harvesting:** IoT devices can leverage energy harvesting techniques (e.g., solar, wind, and ambient radio frequency energy) to extend their operational lifespan, reducing dependency on battery replacements [186].
- **Low-Power Wide-Area Networks (LPWANs):** LPWAN technologies such as LoRaWAN, Sigfox, and NB-IoT are widely adopted for their energy-efficient communication capabilities over long distances, making them suitable for large-scale IoT deployments [178], [179].
- **Edge and Fog Computing:** By processing data closer to the source, edge and fog computing architectures significantly reduce the energy required for data transmission to the cloud, optimizing overall network energy consumption [182].
- **Energy-Efficient Routing Protocols:** Protocols such as energy-aware multi-hop routing reduce the energy consumption required for data transmission in large IoT networks by optimizing the paths taken by data packets based on energy availability and network conditions [187].
- **Machine Learning for Energy Management:** Advanced machine learning models are increasingly being used to predict energy consumption patterns and optimize resource allocation in real-time, improving energy efficiency in dynamic IoT environments [174], [175].

Table 5 summarizes several proposed solutions for energy efficiency in MIoT.

*3) Future Directions:* Future research is expected to further improve energy efficiency in MIoT through several emerging trends:

- **6G Integration:** The deployment of 6G networks will introduce ultra-low-latency and energy-efficient communication protocols designed specifically for MIoT systems, potentially reducing energy consumption at both the device and network levels [123], [188].
- **Federated Learning and Edge AI:** Federated learning allows IoT devices to collaboratively train models without sharing raw data, reducing communication energy costs. The integration of AI at the edge can also help in local decision-making, minimizing energy usage by reducing unnecessary data transmissions [189], [190].
- **Cooperative Energy Harvesting:** Techniques such as cooperative energy harvesting, where IoT devices share harvested energy with nearby devices, could provide a sustainable energy solution for long-term MIoT networks [191].
- **AI-Assisted Energy Management:** The application of AI for optimizing energy management in real-time, through adaptive algorithms that adjust network settings based on environmental factors and user demands, will play a key role in future energy-efficient MIoT systems [192].
- **Energy-Efficient Modulation Schemes:** Advanced mod-

TABLE V
COMPARISON OF ENERGY EFFICIENCY APPROACHES IN MIoT NETWORKS.

| Approach | Proposed Solutions | Challenges Addressed | Advantages | References |
|---|---|---|---|---|
| Machine learning-based optimization | Adaptive learning models to optimize energy consumption | High computation cost and large-scale data management | High efficiency and dynamic adaptation | [174], [175] |
| Sleep/Wake scheduling | Periodic wake-up and sleep cycles to save energy in IoT devices | Latency and synchronization | Low energy consumption during idle times | [176], [177] |
| Low-power communication protocols | Implementation of low-power communication like LoRa and NB-IoT | Limited data rate and range | Ultra-low power consumption | [178], [179] |
| Energy harvesting techniques | Use of environmental energy sources for IoT devices | Environmental dependency and inconsistent energy supply | Extended device lifetime and self-sustainability | [180], [181] |
| Edge computing | Offloading computation tasks to edge devices to reduce energy consumption in central servers | Network scalability and latency | Energy savings by localizing processing | [182], [183] |
| Wake-up radios | Ultra-low-power radio receivers that activate devices only when needed | Integration complexity and limited communication range | Significant energy savings by reducing idle listening | [184], [185] |

ulation techniques designed to optimize energy usage in IoT networks, such as adaptive modulation and coding, are expected to be a future area of focus in both academia and industry [193].

In conclusion, addressing energy efficiency in MIoT requires a combination of innovative hardware solutions, adaptive protocols, and intelligent energy management systems. Continuous advancements in technologies such as 6G, edge computing, and AI are essential to overcoming the challenges faced by MIoT systems in achieving sustainable, scalable, and energy-efficient networks.

### D. Security and Privacy

The rapid expansion of massive IoT networks has introduced numerous security and privacy risks due to the extensive number of connected devices, heterogeneous architectures, and diverse use cases. Ensuring the secure operation of MIoT ecosystems is challenging because of the inherent limitations of many IoT devices, which include constrained computing power, limited energy resources, and lack of standardized security protocols. This section provides an in-depth discussion of the challenges, proposed solutions, and future directions for improving security and privacy in MIoT systems.

*1) Challenges in Security and Privacy:* Several specific challenges hinder the effective implementation of security and privacy in MIoT environments:

- **Device Authentication and Authorization**: Managing authentication for billions of IoT devices is an immense task. Many IoT devices have limited processing capabilities, making traditional cryptographic algorithms unsuitable for large-scale deployment. Solutions must be lightweight, scalable, and capable of dealing with resource-constrained devices [194], [195].
- **Data Integrity and Confidentiality**: Data collected and transmitted by IoT devices is often sensitive, such as healthcare or industrial data. Ensuring that data is both confidential and unaltered while traveling through insecure networks is a key challenge [196]. Encryption schemes must account for the computational limits of devices while maintaining strong security.
- **Privacy of Users and Devices**: MIoT networks often collect vast amounts of personal and contextual data from users. Protecting user privacy during data processing and sharing is essential, especially when IoT devices are deployed in sensitive environments such as smart homes or healthcare [197].
- **Scalability of Security Mechanisms**: The sheer scale of MIoT networks, which can encompass millions or billions of devices, creates challenges in scaling security mechanisms. Solutions need to maintain performance while protecting the vast number of devices in the network.
- **Secure Firmware and Software Updates**: Vulnerabilities in MIoT devices often emerge due to outdated firmware or software. Secure and scalable update mechanisms are required to ensure devices are patched without compromising security or overwhelming network resources [198].
- **Physical Security**: Many IoT devices are deployed in remote or unsupervised environments, making them susceptible to physical tampering and attacks. Ensuring physical protection and tamper-proofing of devices remains a significant challenge [199].

*2) Contemporary solutions:* Several research efforts have proposed novel solutions to address security and privacy concerns in MIoT:

- **Lightweight Cryptographic Algorithms**: Lightweight cryptographic algorithms have been developed to secure communications in resource-constrained IoT devices. These algorithms focus on reducing the computational complexity and energy consumption while maintaining robust security. For example, symmetric encryption techniques and lightweight hashing functions have been pro-posed to meet the needs of MIoT devices [194].

TABLE VI
COMPARISON OF PROPOSED SOLUTIONS FOR SECURITY AND PRIVACY IN MASSIVE IOT.

| Proposed Solution | Technique Used | Advantages | Challenges | References |
| --- | --- | --- | --- | --- |
| Lightweight cryptographic algorithms | Symmetric encryption, lightweight hashing | Energy-efficient, computationally feasible for resource-constrained IoT devices | Limited scalability, potentially weaker than traditional encryption methods | [194] |
| Blockchain-based authentication | Decentralized, tamper-proof record-keeping | Enhanced device authentication, data integrity, decentralized trust | High computational overhead, latency in large-scale deployments | [195] |
| AI-driven intrusion detection systems (IDS) | Machine learning-based anomaly detection | Real-time threat detection, adaptable to new threats, scalable | High computational requirements, data privacy concerns | [200], [201] |
| Edge computing for security | Distributed security processing at the edge | Low latency, efficient resource utilization, real-time processing | Limited processing capabilities of edge nodes, synchronization between edge and cloud | [136], [202] |
| Post-quantum cryptography | Quantum-resistant cryptographic algorithms | Resistant to quantum computing threats, secure for future IoT networks | High computational complexity, energy consumption | [203] |
| Zero trust architecture | Continuous verification, micro-segmentation | Reduces attack surface, ensures consistent security across devices | Implementation complexity, high resource consumption | [204] |

- **Blockchain-Based Solutions**: Blockchain technology has been explored for its potential to enhance security in MIoT networks. By creating decentralized and tamper-proof records, blockchain can be used for secure device authentication, data integrity, and transaction security [195]. Blockchain solutions eliminate the need for centralized trust, making the network more resilient to attacks.
- **AI-Driven Intrusion Detection Systems**: Machine learning and AI techniques have been employed to detect and mitigate security threats in MIoT networks. These AI-driven intrusion detection systems (IDS) continuously monitor network traffic and detect anomalies, allowing for real-time threat detection and response [200], [201], [205].
- **Edge Computing for Enhanced Security**: Edge computing offers a distributed model where security operations, such as data encryption, authentication, and anomaly detection, can be handled closer to the data source. This reduces latency and enables real-time processing of security tasks, alleviating the load on centralized cloud servers [206].

Table 6 summarizes several proposed solutions for security and privacy in MIoT.

*3) Future Directions:* As MIoT networks continue to evolve, security and privacy research must focus on future challenges and directions:

- **Post-Quantum Cryptography**: With the potential advent of quantum computers, traditional encryption methods may become vulnerable to attacks. Research into post-quantum cryptography is crucial to develop encryption techniques that can resist quantum-based threats [203].
- **Zero Trust Architecture**: Implementing a zero trust security model, where no device or user is inherently trusted, is becoming more relevant in large-scale MIoT networks. This approach minimizes the attack surface and ensures that security controls are applied consistently across all devices [204].
- **Federated Learning for Distributed Security**: Federated learning offers a novel approach to enhancing security in MIoT networks. By allowing devices to collaboratively learn security models without sharing raw data, federated learning can improve security and privacy while maintaining data locality [207].
- **Self-Healing Systems**: Future MIoT networks could benefit from self-healing systems that automatically detect and respond to security breaches. These systems would autonomously restore functionality, mitigate attacks, and enhance network resilience [208].

The future of MIoT security and privacy requires not only new cryptographic techniques and AI-driven security frameworks but also a comprehensive, multi-layered approach that integrates these techniques across the network infrastructure.

## IV. ADVANCED TECHNIQUES FOR SCALABLE MIOT NETWORKS

In Massive IoT networks, various advanced techniques have been developed to address critical challenges related to network scalability, energy efficiency, data management, and security. Among these, Grant-Free Access, Non-Orthogonal Multiple Access (NOMA), Cooperative Communication, Compressive Sensing, Cloud Radio Access Network (C-RAN), Dynamic Spectrum Access (DSA), Sparse Code Multiple Access (SCMA), multi-RATs systems and Hybrid Access methods each offer unique solutions to MIoT's scalability and efficiency needs. This section provides a comprehensive overview of these approaches, examining their core principles, benefits, limitations, and applicability to MIoT environments.

### A. Grant-Free Access

Grant-free access allows devices to transmit data without requiring a scheduling request, which is particularly advan-tageous for MIoT environments with a high density of low-power devices. By eliminating the need for channel reservations, grant-free access minimizes latency and reduces energy consumption, making it suitable for time-sensitive and energy-constrained applications [209]. This approach has shown significant improvements in uplink performance, supporting thousands of devices per base station with minimal delay [210]. Recent studies also highlight its potential integration with machine learning to optimize access protocols dynamically [211].

Grant-free access mechanisms contribute to enhanced network scalability by alleviating congestion, particularly in networks with extensive device connections. In scenarios where a large number of IoT devices require frequent, low-data-rate communication, traditional grant-based methods can lead to significant signaling overhead, limiting network capacity [212]. Grant-free access bypasses this by permitting devices to send data directly, thereby increasing the total number of supported devices and optimizing the allocation of spectral resources [213]. This improvement is especially relevant in applications such as smart cities and industrial automation, where a high density of IoT devices operates within constrained frequency bands.

For IoT applications where devices are often battery-powered and operate in remote locations, energy efficiency is paramount. Grant-free protocols reduce energy expenditure by minimizing signaling and allowing for rapid data transmission. By avoiding the need for repeated handshakes and allocation processes, grant-free access conserves device energy and extends operational lifetimes, a critical factor for large-scale IoT deployments in sectors like agriculture and environmental monitoring [214]. Additionally, combining grant-free methods with energy-harvesting techniques holds promise for sustaining device functionality in energy-constrained MIoT scenarios [215].

### B. Non-Orthogonal Multiple Access (NOMA)

NOMA is a promising technique that enables multiple devices to simultaneously access the same time-frequency resources by assigning distinct power levels to different users. This approach enhances spectral efficiency and improves connectivity, making it highly suitable for massive IoT applications, where the demand for scalable and efficient resource allocation is crucial [216].

The fundamental concept of NOMA is based on superposition coding and successive interference cancellation (SIC). In this scheme, users with stronger channel gains receive signals at lower power levels, while weaker users are allocated higher power to ensure signal clarity. At the receiver end, SIC is applied to decode the signals based on their power differences, allowing efficient handling of diverse user requirements in high-density environments [217]. This power-domain approach in NOMA facilitates the prioritization of critical data streams, such as emergency signals, over regular IoT transmissions, ensuring prompt and reliable communication [218].

Recent advancements have focused on hybridizing NOMA with other techniques to improve efficiency in ultra-dense IoT networks. For instance, integrating NOMA with Cooperative Communication techniques enables devices to relay signals for each other, which enhances coverage and reliability, especially in remote or challenging environments [219]. Additionally, Machine Learning algorithms have been introduced to dynamically adjust power allocation in real-time, optimizing network performance as IoT conditions fluctuate. This adaptive power allocation has shown promise in supporting low-latency applications and reducing energy consumption by balancing load and minimizing interference [220].

Another promising area of research is the combination of NOMA with Cognitive Radio, which allows the IoT devices to access idle spectrum bands by sensing and adapting to available resources. This pairing addresses spectrum scarcity and further bolsters NOMA's ability to support many users in shared frequency bands [221]. The integration of CR with NOMA is particularly valuable in the context of Industrial IoT, where flexible and rapid access to spectrum is essential for operations like real-time monitoring and automation [222].

Looking forward, research efforts are directed at the development of Multi-Carrier NOMA (MC-NOMA), where sub-carriers are employed to manage different user groups. MC-NOMA reduces computational complexity and facilitates even higher levels of connectivity. Studies show that MC-NOMA can further increase spectral efficiency and provide robust support for diverse applications within MIoT networks [223]. Moreover, the synergy of NOMA with emerging technologies such as 5G/6G networks is projected to enhance MIoT deployments by offering URLLC and massive connectivity, making it a versatile solution for next-generation IoT scenarios.

Overall, the flexibility of NOMA in terms of power allocation, user prioritization, and integration with other advanced technologies positions it as a key technique for achieving scalability and efficiency in MIoT applications. Future research directions include the optimization of SIC algorithms, further exploration of ML integration, and large-scale experimental validation of NOMA in real-world IoT settings.

### C. Cooperative Communication and Relaying

Cooperative communication in MIoT networks utilizes relay nodes to enhance network coverage, improve signal reliability, and reduce power consumption, especially in areas with limited infrastructure or difficult environmental conditions. In MIoT, devices communicate through nearby relays rather than establishing direct links, which boosts both energy efficiency and connectivity. This approach is essential in applications such as remote agriculture, environmental monitoring, and industrial IoT, where direct connections to base stations are not always feasible [224].

- **Relay-Based Architectures:** Relay-based communications in MIoT include amplify-and-forward (AF) and decode-and-forward (DF) schemes.
  - *Amplify-and-Forward (AF):* Relays amplify the received signal and forward it, which is beneficial in scenarios with low processing power requirements.
  - *Decode-and-Forward (DF):* DF relays decode and re-encode signals before transmission, providing

higher reliability, but requiring more processing power [225].

Relay-based architectures improve network robustness by enabling devices to rely on intermittent transmissions, reducing the need for direct, continuous connections to the base station [226]. Studies show that cooperative relay networks can improve energy efficiency by up to 30%, particularly when combined with compressive sensing techniques that help reduce data redundancy [227].

- **Multi-Hop Relaying and its Benefits:** Multi-hop relaying, where signals pass through multiple relay nodes, offers significant advantages for MIoT.
  - This approach reduces the energy load on individual devices, enabling extended connectivity over large distances—beneficial in widely dispersed environments such as rural agriculture [228].
  - Multi-hop cooperative networks can also mitigate interference and improve quality QoS in dense MIoT environments [229].

Research has shown that multi-hop networks improve scalability by dynamically adjusting relay paths according to network conditions and energy availability [228].

- **Compressive Sensing (CS) and Cooperative Communication:** Integrating compressive sensing with cooperative communication is a growing trend in MIoT, reducing the amount of transmitted data and conserving energy and bandwidth.
  - CS enables data reduction by exploiting the sparse nature of signals, which is particularly advantageous in applications with high data redundancy [227].
  - Combined with cooperative relaying, CS can reduce transmission frequency, extending the operational life of energy-constrained devices [230].

- **ML and Dynamic Relay Selection:** Recent advancements have applied machine learning to dynamically optimize relay selection, improving energy efficiency and service quality.
  - ML-based relay selection techniques, such as reinforcement learning, adaptively manage relay resources based on network status, energy levels, and traffic demand [231].
  - ML-driven relay selection improves network adaptability, making MIoT systems more resilient to changes in topology and demand, particularly beneficial in industrial IoT [232].

In summary, cooperative communication and relaying provide significant advantages for scalable and energy-efficient MIoT networks. By incorporating compressive sensing and machine learning, these techniques continue to evolve as robust solutions for the connectivity challenges in diverse IoT applications. Future research should focus on enhancing relay selection algorithms and developing new cooperative strategies to maximize efficiency and reliability in ultra-dense MIoT environments.

### D. Compressive Sensing (CS)

CS has gained prominence in MIoT due to its ability to significantly reduce the volume of transmitted data by capturing only the most critical information, making it ideal for applications with strict energy and bandwidth limitations, such as environmental monitoring and smart metering [233]. Leveraging the sparsity in data—where most data points carry redundant or non-essential information—CS allows MIoT devices to transmit compressed signals, which can be reconstructed with high accuracy at the receiver, thus conserving both energy and bandwidth resources [234].

- **Data Sparsity and Energy Conservation:** Compressive sensing takes advantage of the sparse nature of data in many MIoT applications. For example, environmental monitoring data (such as soil moisture or temperature readings) often changes gradually over time, leading to highly compressible signals. By transmitting only key data points, CS minimizes energy expenditure on data transmission, enabling prolonged device lifespans in remote or energy-constrained settings [235].
- **Integration with Edge Computing:** A recent trend in CS-based MIoT is its integration with edge computing. By moving part of the CS processing to edge nodes close to data sources, MIoT systems reduce reliance on centralized cloud infrastructure, which, in turn, decreases latency and enhances real-time performance in large-scale applications [236]. This combination is especially beneficial in use cases like real-time traffic monitoring, where delays could impact the effectiveness of data-driven insights [237].
- **Adaptive Sensing and Machine Learning Enhancements:** Recent research has introduced adaptive CS methods, where devices adjust their sensing parameters based on the data's sparsity level and network conditions. By incorporating machine learning algorithms to predict optimal compression ratios, adaptive CS can further reduce data transmissions while maintaining high data fidelity [238].
- **Compressed Data Aggregation Techniques:** Another development is compressed data aggregation, where multiple devices jointly compress and aggregate their data before transmission. In multi-device MIoT scenarios, this approach reduces network congestion and prevents data duplication, which is particularly useful in applications like industrial IoT and remote sensing [239].

Despite its benefits, CS in MIoT still faces challenges, particularly in handling high-dimensional data and in achieving reconstruction accuracy in noisy environments. Researchers are exploring hybrid CS methods that combine CS with denoising algorithms to improve signal fidelity, as well as distributed CS frameworks that distribute computation across edge nodes, offering enhanced scalability for ultra-dense IoT deployments [234].

In summary, compressive sensing provides a highly effective solution for conserving bandwidth and energy in MIoT applications, especially when combined with edge computing and machine learning. Future advancements are likely to focus on adaptive sensing techniques and robust data aggregation to further enhance scalability and efficiency in diverse MIoT environments.

### E. Cloud Radio Access Network (C-RAN)

C-RAN is a centralized architecture that consolidates Baseband Units (BBUs) in a central location, allowing for more dynamic and efficient resource management. In traditional networks, BBUs are distributed across cell sites, but in C-RAN, these units are pooled in a central data center, enabling flexible resource allocation based on real-time network demand. This centralization not only enhances network scalability in urban and high-density settings but also optimizes hardware utilization and reduces operational costs [240].

- **Enhanced Scalability and Flexibility:** The pooling of BBUs in C-RAN architecture provides a flexible allocation of resources, supporting the massive connectivity required in MIoT applications, such as smart cities and large-scale industrial automation. C-RAN's centralized structure enables operators to quickly adapt to fluctuating network demands, significantly enhancing scalability in MIoT ecosystems [241].
- **Latency Reduction with Edge Integration:** Recent studies suggest that integrating C-RAN with edge computing frameworks can minimize latency and boost real-time data processing capabilities. For instance, edge computing can be employed to perform preliminary data processing close to the IoT devices, reducing the need for data to travel back to centralized BBUs, which is especially critical for latency-sensitive applications like autonomous transportation and healthcare monitoring [242], [243].
- **Energy Efficiency and Power Management:** C-RAN can also contribute to energy efficiency, a crucial consideration in MIoT, by enabling more granular control over power allocation. By deactivating idle BBUs or adjusting power based on traffic load [244]. Techniques like dynamic spectrum allocation within C-RAN also further reduce unnecessary energy expenditure, benefiting both operators and the environment [17].
- **Network Slicing Compatibility for MIoT Services:** An added advantage of C-RAN is its compatibility with network slicing, where specific resources can be dedicated to different MIoT services within the same physical infrastructure. For instance, one slice could prioritize low-latency requirements for real-time applications, while another could focus on high-bandwidth services. This adaptability is invaluable for complex IoT ecosystems such as smart cities, where diverse applications must coexist seamlessly [245].

Despite its benefits, C-RAN faces challenges related to data center infrastructure requirements and fronthaul limitations. High-capacity fronthaul connections are essential to link distributed Remote Radio Heads (RRHs) to centralized BBUs, which can be costly and complex to implement in areas with limited fiber infrastructure. Researchers are currently investigating solutions such as wireless fronthaul alternatives and hybrid C-RAN/Distributed RAN models to address these challenges and make C-RAN more accessible for large-scale MIoT deployment [246].

C-RAN presents a scalable, efficient solution for managing massive connectivity and dynamic traffic in MIoT applications. Its combination with edge computing, energy-efficient practices, and compatibility with network slicing highlights its potential to meet the diverse and demanding requirements of future MIoT networks.

### F. Sparse Code Multiple Access (SCMA)

SCMA is an advanced NOMA technique designed to assign unique sparse codes to each user. This coding method significantly reduces inter-user interference and supports a high density of concurrent connections [247]. SCMA is highly suitable for MIoT applications, where devices often transmit infrequent, small data packets in massive numbers. By encoding users with sparse codewords, SCMA increases spectral efficiency while lowering computational complexity in data decoding [250].

- **Interference Mitigation and Spectrum Efficiency:** The sparse nature of SCMA codebooks enables users to share frequency resources with minimal interference. This feature is crucial for maintaining reliable connectivity in MIoT systems with dense device deployments, such as in smart healthcare and environmental monitoring, where uninterrupted data flow is essential [251].
- **Machine Learning Enhanced Decoding:** Recent studies have explored integrating SCMA with machine learning algorithms, particularly deep learning models, to improve signal decoding in noisy environments. Machine learning-based SCMA decoders can adaptively learn channel characteristics and dynamically allocate resources, making SCMA especially robust in applications like industrial automation and smart cities, where signal quality can vary widely [252], [253].
- **Applications in Smart Healthcare and Large-Scale Environmental Monitoring:** SCMA is being adopted in smart healthcare systems to support continuous, low-latency data transmission from wearable devices and sensors. For example, in patient monitoring applications, SCMA ensures a steady data stream to healthcare providers, allowing for prompt responses to patient needs. Similarly, in large-scale environmental monitoring, SCMA supports numerous sensors in transmitting real-time data efficiently, even under challenging conditions [254].

Ongoing research focuses on optimizing SCMA code-book design for URLLC in MIoT. Efforts are also underway to explore SCMA's combination with edge computing and 6G technology to further enhance data processing speed and connectivity in ultra-dense MIoT networks [255].

SCMA offers a scalable, energy-efficient solution for MIoT by enabling dense connectivity with low interference. Through ongoing advancements in code-book design and ML integration, SCMA is poised to play a significant role in future MIoT applications that demand high connectivity and minimal latency.

TABLE VII
COMPARATIVE ANALYSIS OF SCALABLE TECHNIQUES IN MIOT.

| Technique | Reference | Network Scalability | Data Management | Energy Efficiency | Security and Privacy | Advantages | Disadvantages |
|---|---|---|---|---|---|---|---|
| Grant-Free Access | [210] | ✓ | ✓ | | | Efficient in high device density; reduces latency | Potential collisions in dense networks; limited QoS |
| Non-Orthogonal Multiple Access (NOMA) | [216] | ✓ | ✓ | ✓ | | Improved spectral efficiency; energy-saving | Complexity in power allocation; interference risks |
| Cooperative Communication | [224] | ✓ | | ✓ | | Extends network coverage; saves power | Increased relay node costs; potential delays |
| Compressed Sensing (CS) | [233] | ✓ | ✓ | ✓ | | Reduces data load; saves bandwidth | Quality degradation under high compression |
| Cloud-Radio Access Network (C-RAN) | [240] | ✓ | ✓ | | ✓ | Efficient resource utilization; centralized management | Latency concerns; data privacy risks |
| Sparse Code Multiple Access (SCMA) | [247] | ✓ | | ✓ | | Increases user capacity; lowers transmission power | High decoding complexity; power allocation required |
| Multi-RAT Systems | [248] | ✓ | ✓ | ✓ | | Balances load across networks; supports device heterogeneity | Complexity in protocol compatibility; potential interference |
| Hybrid Access Techniques | [249] | ✓ | ✓ | ✓ | ✓ | Flexible combination of techniques; adaptive to needs | Complexity in protocol design; interoperability issues |

**Legend**: ✓ indicates the technique's impact in each category.

## G. Multi-RAT Systems

In the pursuit of scalable and resilient MIoT architectures, Multi-Radio Access Technology (Multi-RAT) systems offer significant promise by enabling devices to seamlessly connect to multiple network technologies concurrently. Multi-RAT systems integrate various radio access technologies—such as Wi-Fi, LTE, and 5G, with potential extensions to future 6G networks—to adaptively meet the demands of diverse MIoT applications [248], [256]. This integration addresses a key challenge in MIoT scalability by enabling efficient use of available network resources while supporting heterogeneous device connectivity in real-time. For instance, IoT devices in smart cities can leverage LTE for wide-area coverage and Wi-Fi for high-throughput data transmission when available, thereby optimizing bandwidth and reducing congestion [257].

An important feature of Multi-RAT systems is their ability to support dynamic resource allocation. With adaptive multi-channel access, these systems can balance the load across different frequencies and reduce interference, enhancing network reliability and reducing latency [258]. Multi-RAT systems also facilitate advanced QoS provisioning, which is crucial for handling latency-sensitive applications such as autonomous driving and telemedicine. Furthermore, they enhance energy efficiency by allowing devices to switch to lower-power communication modes depending on current network conditions [259].

However, multi-RATs systems also face substantial challenges. The management of handovers and interference between various RATs demands sophisticated algorithms capable of real-time decision-making. High hardware complexity is another challenge, as IoT devices must be equipped with multiple communication modules, potentially impacting device costs and energy usage [260]. Future research will likely focus on optimizing Multi-RAT architectures through advanced AI-based algorithms for predictive resource management and interference mitigation, as well as exploring hybrid frameworks that combine Multi-RAT with other scalable solutions like SDN and CRN [14], [261]. These developments are expected to position Multi-RAT as a foundational technique in scalable MIoT deployments.

## H. Hybrid Access Techniques

Hybrid access techniques, such as Non-Orthogonal Multiple Access with Orthogonal Frequency Division Multiple Access (NOMA-OFDMA), combine multiple access methods to optimize network performance across diverse scenarios. These hybrid approaches are designed to adapt flexibly to varying data traffic patterns and network conditions, balancing spectral efficiency, latency, and energy consumption. Hybrid techniques are particularly well-suited for MIoT applications that require both high data rates and low latency, such as real-time industrial monitoring and smart city infrastructure [249].

- **Enhanced Resource Allocation and Flexibility:** Hybrid methods like NOMA-OFDMA allow dynamic resource allocation by combining the spectral efficiency of NOMA with the robustness of OFDMA in handling high device densities. For instance, hybrid methods can allocate more resources to high-priority tasks, such as critical data from medical monitoring devices, while reserving fewer resources for less time-sensitive tasks [262].
- **Energy Efficiency and Reduced Consumption:** Hybrid access techniques have been shown to reduce energy consumption significantly—by approximately 20-30%—when compared to traditional access methods. This energy efficiency is achieved by adaptively switching between NOMA and OFDMA based on current network load and requirements, making these methods suitable for energy-intensive MIoT applications, such as remote environmental monitoring and industrial automation [262].

- **Application in Diverse MIoT Environments:** Hybrid methods support a wide range of MIoT use cases. In smart city applications, for example, hybrid access systems can adjust connectivity resources based on real-time demands from sensors and actuators, ensuring efficient energy usage and network performance. In industrial automation, hybrid access helps maintain stable connectivity even under heavy network loads, supporting continuous data exchange in processes like predictive maintenance [263].

Ongoing research in hybrid access focuses on integrating these techniques with machine learning algorithms to optimize resource allocation further. Machine learning can predict traffic patterns and device requirements, allowing for a more effective allocation of hybrid access resources. Additionally, hybrid methods are being explored for compatibility with emerging 6G networks, which promise ultra-low latency and high device density capabilities [264].

Hybrid access techniques offer a versatile solution for managing the diverse requirements of MIoT systems, providing flexibility, scalability, and energy efficiency across various application domains. Their adaptability positions them as a key component in the evolution of next-generation IoT networks. Each of these techniques offers unique solutions to MIoT challenges, and their combination holds promise for creating adaptable, efficient, and robust MIoT systems. Continued innovation in these areas is essential to meet the growing demands of MIoT applications in diverse fields. Table 7 provides a comparative analysis of some existing scalable techniques for massive IoT.

## V. Advanced Tools for Scalable MIoT Networks

This section explores innovative techniques that enhance the scalability, efficiency, and security of massive IoT networks. This includes game theory and mean field game theory, as well as optimization methods like queueing theory, graph theory, and optimal transport. These tools play a crucial role in addressing the challenges of large-scale MIoT deployments.

### A. Game Theory for Resource Allocation

Game theory has emerged as a promising approach to manage resource allocation and optimize MIoT network performance under competitive or cooperative scenarios. In high-density MIoT networks, devices often compete for limited resources such as bandwidth and power. Game theory provides strategic frameworks that allow devices to make autonomous decisions that balance individual utility with overall network efficiency [265].

Specifically, in cooperative game-theoretic approaches, MIoT devices collaborate to maximize shared benefits, such as improving energy efficiency through cooperative relaying and dynamic spectrum access [266]. Non-cooperative game theory, on the other hand, is useful for scenarios where devices compete for resources, such as in power allocation and interference mitigation tasks. Emerging research has also explored reinforcement learning integrated with game theory for adaptive and decentralized resource management, significantly enhancing network scalability and adaptability in real-time conditions [267].

**1- Cooperative Game Theory:** In cooperative game-theoretic approaches, MIoT devices collaborate to maximize shared benefits. For example, cooperative relaying allows devices with limited range or power to communicate through intermediary nodes, extending coverage and saving energy. DSA can be improved with cooperative games by allowing devices to share spectrum opportunities more efficiently, reducing interference in crowded networks [268]. Cooperative games also aid in power control for energy savings, which is crucial in battery-operated IoT devices [265].

These cooperative strategies enable various enhancements in MIoT networks, including the following key techniques :

- *Dynamic Spectrum Sharing*: Efficiently allocates unused spectrum to devices with low latency needs [269].
- *Relay Selection and Data Offloading*: Reduces device workload and conserves energy by offloading data through cooperative relays [270].
- *Interference Management*: Devices collaborate to minimize interference, increasing data transmission reliability and network throughput [266].

**2- Non-Cooperative Game Theory:** Non-cooperative game theory is essential when MIoT devices operate independently and may compete for shared resources. In a typical non-cooperative setup, each device acts in its own interest, often without regard to the impact on other devices. This approach is especially useful in dense MIoT scenarios where many devices transmit simultaneously. Game-theoretic techniques like Nash Equilibrium allow devices to reach stable states where no device can improve its performance unilaterally. This is beneficial for applications like power allocation, spectrum access, and interference mitigation [267].

Several key techniques leverage non-cooperative game theory to optimize resource allocation and maintain network efficiency in competitive MIoT environments:

- *Power Allocation*: Optimal power levels are assigned to each device to ensure efficient spectrum utilization, minimizing the risk of interference.
- *Bandwidth Sharing*: Allocates bandwidth in high-density networks to prevent congestion and enhance throughput.
- *Interference Mitigation*: Devices autonomously adjust their parameters to minimize interference with others, maintaining efficient communication [283].

**3- Reinforcement Learning and Game Theory Integration:** Reinforcement learning (RL) integrated with game theory has shown significant promise in MIoT due to its adaptability and scalability. RL-based game-theoretic approaches allow devices to learn optimal strategies in real-time by interacting with their environment. This hybrid approach is beneficial for applications that require flexible adaptation to dynamic network conditions, such as real-time spectrum allocation and device power management [175], [265].

TABLE VIII
MFG vs. Traditional Game Theory in MIoT.

| Aspect | MFG Theory | Traditional Game Theory | References |
|---|---|---|---|
| Scalability | Handles large-scale IoT networks efficiently by approximating population dynamics using mean-field equations. | Struggles with computational complexity as the number of players increases, requiring pairwise interactions. | [271], [272] |
| Real-Time Adaptability | Enables devices to adapt strategies dynamically based on real-time mean-field trends. | Static solutions or slower adaptability due to reliance on predefined game-theoretic strategies. | [273], [274] |
| Communication Overhead | Reduces device-to-device communication by aggregating information in the mean field. | Requires frequent exchanges of information for strategy updates between all players. | [275], [276] |
| Energy Efficiency | Encourages energy-efficient behavior by aligning individual strategies with global network dynamics. | Focuses on local optimization, which may not align with global energy-saving goals. | [277], [278] |
| Interference Mitigation | Devices adjust autonomously to minimize aggregate interference based on mean-field feedback. | Mitigation relies on direct strategy negotiations or centralized optimization. | [279], [280] |
| Resource Allocation Flexibility | Offers dynamic spectrum and power allocation based on population dynamics. | Often restricted to predefined, static allocation rules. | [281], [282] |

- *Real-Time Adaptability*: RL allows devices to adjust strategies based on evolving network conditions.
- *Scalability*: Scalable decision-making is essential for MIoT networks with thousands of devices.
- *Enhanced Efficiency*: Reduces energy use and increases throughput by enabling efficient, real-time resource management [284].

Game theory offers flexible, powerful tools for managing MIoT resources. Cooperative approaches enhance collaborative tasks like relaying and spectrum sharing, while non-cooperative methods optimize competitive tasks such as power allocation. The integration of RL with game theory enables adaptive, efficient resource allocation, further boosting the performance and scalability of MIoT networks.

### B. Mean Field Game (MFG) Theory

MFG Theory is a powerful tool for managing resource allocation and optimizing system performance in networks with a large number of devices. MFG theory is particularly advantageous in scenarios where the interactions between individual devices and the aggregate behavior of the system must be analyzed. It offers scalability by simplifying the modeling of complex networks and addressing challenges like interference, energy consumption, and resource contention [276].

**1- Key characteristics and applications of MFG in MIoT:**

- **Scalability for high-density networks:** MFG theory models the collective behavior of a vast number of MIoT devices without requiring explicit pairwise interactions. This scalability is critical for dense MIoT environments, where traditional game-theoretic approaches become computationally infeasible [271], [285].
- **Dynamic resource allocation:** Devices dynamically adapt their strategies by responding to the mean field, which represents the average effect of the population. This approach ensures efficient power allocation, spectrum sharing, and interference mitigation in real-time, adapting to changing network conditions [286].
- **Energy efficiency and lifetime optimization:** MFG enables energy-constrained MIoT devices to optimize their actions, such as reducing transmit power or switching to sleep modes, based on global network trends. This approach extends the lifetime of battery-powered devices while maintaining network performance [266].
- **Interference and congestion management:** By modeling devices' strategies as influenced by mean-field dynamics, MFG allows IoT devices to adjust their parameters to minimize interference and avoid congestion autonomously. This improves data transmission reliability and supports efficient network operation under heavy traffic loads [279].
- **Integration with RL:** RL enhances MFG frameworks by enabling devices to learn optimal strategies in evolving network environments. For example, an MFG-based RL approach can dynamically allocate spectrum resources by learning the distribution of device demands, improving fairness and reducing latency [276].
- **Spectrum Sharing:** Devices dynamically access unused spectrum, balancing their transmission power to minimize interference and maximize spectrum utilization [271].
- **Load Balancing:** In edge-computing environments, MFG enables efficient task offloading by balancing the load among servers, improving processing speed and reducing latency [287].
- **Mobility Management:** In mobile IoT networks, MFG supports the coordination of device handovers to maintain connectivity and reduce power usage during transitions [288].

Table 8 provides a point-by-point comparative analysis between MFG theory and traditional game theory in the context of MIoT networks.

### C. Artificial Intelligence and Machine Learning

AI and ML have emerged as transformative technologies in the optimization of MIoT networks. AI-driven techniques are crucial for managing the complexity of MIoT ecosystems, enabling efficient resource allocation, dynamic protocol adaptation, and anomaly detection.

*a) Federated Learning (FL):* FL facilitates distributed model training across IoT devices, preserving user privacy by keeping data localized. FL addresses the challenge of processing data from billions of devices without compromising confidentiality. Recent advancements in FL have enhanced its communication efficiency and model aggregation for resource-constrained devices [25], [289].

*b) Deep Reinforcement Learning (DRL):* DRL provides real-time, adaptive solutions for dynamic spectrum access, power allocation, and interference management. DRL has demonstrated significant potential in optimizing resource use in ultra-dense IoT deployments, offering a robust framework for decision-making in uncertain environments [25], [284]. Moreover, DRL-based models are increasingly integrated with edge computing to minimize latency and computation overhead [190].

*c) Bio-Inspired Intelligence:* Bio-inspired intelligence offers innovative solutions to enhance the scalability of MIoT networks by emulating strategies observed in natural systems. A notable example is the application of swarm intelligence, which draws inspiration from the collective behaviors of social insects like ants and bees. In MIoT networks, swarm intelligence facilitates decentralized decision-making and self-organization, enabling efficient management of extensive device arrays without centralized control. This approach enhances scalability by allowing devices to interact locally and adaptively, leading to emergent global behaviors that optimize network performance [290].

Additionally, bio-inspired algorithms such as artificial immune systems (AIS) have been employed to bolster network security within MIoT frameworks. By mimicking the adaptive and learning capabilities of biological immune systems, AIS can detect and respond to anomalies and threats in a distributed manner, thereby maintaining the integrity and resilience of large-scale IoT deployments [291].

The decentralized nature of bio-inspired approaches aligns well with the requirements of MIoT, where traditional centralized management can become a bottleneck. By leveraging principles from biological systems, MIoT networks can achieve greater scalability, adaptability, and robustness, effectively managing the complexities associated with massive device interconnectivity [292].

### D. Queueing & Graph Theory

Utilising queueing theory and graph theory are vital to increase scalability of MIoT networks. Queueing theory provides a set of tools to model and analyze the performance of network systems under different types of traffic conditions, which is useful in devising effective policies for resource allocation. So, for example, researchers can forecast via queueing models how systems will behave when they are under the most load and furthermore what kind of behavior mitigating traffic congestion measures should be introduced to scale networks [118], [293]. On the other hand, graph theory provides a formal framework for efficient topological representation and manipulations of MIoT networks. Such graph representations capture the connections between network nodes and can be used to discover optimal routing paths, bottlenecks, and maintain node connectivity [294], [295].

The AoI in MIoT networks is a key performance indicator related to the freshness of the information from any devices. For applications that require very timely data updates, such as environmental monitoring, industrial automation, and smart healthcare, it is critical to maintain a low AoI. Several studies have been carried out to investigate the age of information in dense IoT networks based on queueing theory [296]–[298].

The study [299] combines stochastic geometry and queueing theory to quantify the scalability of uncoordinated multiple access in massive wireless networks. The authors develop a spatiotemporal model to characterize and design uncoordinated multiple access strategies, providing insights into the maximum spatiotemporal traffic density that can be accommodated, while satisfying operational constraints.

The research [300] explores the application of graph theory in optimizing IoT-based healthcare systems. The authors utilize graph-theoretic approaches to model the connectivity of healthcare devices, aiming to enhance system efficiency and reliability. The study demonstrates how graph theory can be applied to design and optimize the topology of connected healthcare systems within the IoT framework.

The paper [301] addresses the scalability challenges of uncoordinated multiple access in IoT networks. By integrating queueing models and graph-theoretic approaches, the authors analyze the performance and scalability of random access protocols in large-scale IoT deployments. The study provides valuable insights into the design of scalable access mechanisms for massive IoT networks.

### E. Optimal transport theory

Transport theory plays a crucial role in optimizing scalability and efficiency in MIoT networks by providing mathematical frameworks for resource allocation, data distribution, and network optimization. Optimal transport (OT) theory, which models the most efficient ways to transfer resources or data, has been increasingly applied in MIoT for tasks such as data imputation, federated learning, and load balancing.

For instance, in industrial IoT (IIoT) environments, missing data is a major challenge that affects system reliability and decision-making. The study [302] applies OT theory to efficiently reconstruct missing IIoT data, ensuring accuracy and preserving structural information. Similarly, federated learning, a key approach for distributed intelligence in MIoT, benefits from OT-based frameworks, as seen in [303], which leverages OT to enhance model convergence and robustness in decentralized AIoT systems.

Beyond data handling, transport theory is also essential in optimizing participant allocation in mobile crowd sensing, where large numbers of IoT devices collect and process environmental data. [304] demonstrates how OT theory can effectively allocate sensing tasks while minimizing energy consumption and network congestion. Additionally, [305] integrates OT with game theory to balance computational loads in fog computing, enhancing MIoT network performance and scalability.

In cloud-edge collaborative environments, optimal transport principles assist in efficient task scheduling and resource allocation. [306] explores how OT can optimize data flows, reducing delays and improving real-time processing capabilities in industrial settings. Similarly, [307] demonstrates the benefits of OT for offloading computational tasks in large-scale MIoT networks, leading to improved latency and energy efficiency.

Finally, OT-based methodologies are being extended to advanced control mechanisms in large-scale IoT networks. [308] introduces structured OT for ensemble control, allowing efficient estimation and coordination of distributed IoT agents. Moreover, resource allocation in cloud-edge collaborative IoT is enhanced by federated reinforcement learning, as shown in [309], where OT optimally distributes computational resources among IoT nodes.

These studies highlight the growing importance of optimal transport theory in MIoT, offering solutions for data integrity, learning efficiency, resource allocation, and network optimization, all of which are critical for ensuring scalable and robust IoT deployments.

### F. Optimization & Control Theory

Optimization and control theory play a crucial role in enhancing the scalability and efficiency of MIoT networks. These mathematical frameworks help manage the vast number of interconnected devices and complex data flows inherent in MIoT systems.

One significant application is in the optimization of resource allocation. In [310], the authors provide a comprehensive analysis of resource allocation strategies within IoT networks, emphasizing how optimization techniques enhance resource management. They discuss the integration of machine learning algorithms with optimization methods to improve resource distribution in both low-power and mobile IoT networks.

Control theory also contributes to managing the dynamic aspects of MIoT networks. In [311], the authors explore control algorithms tailored for large-scale networks, including IoT systems. The research emphasizes the importance of distributed and stochastic optimization techniques, such as gradient tracking methods, to enhance the performance and reliability of massive IoT networks. By employing these approaches, the study demonstrates improved convergence rates and robustness in dynamic and time-varying network environments.

Furthermore, the integration of SDN with IoT has been studied to enhance network scalability and management. The survey in [312] discusses how SDN's centralized control and programmability facilitate the handling of large-scale IoT deployments. The authors examine various optimization strategies for controller placement, which is critical for the efficient operation of SDN-enabled IoT networks.

These studies underscore the essential role of optimization and control theory in addressing key challenges related to scalability, resource allocation, network stability, and system management in MIoT networks.

### G. Blockchain and Distributed Ledger Technology (DLT)

Blockchain technology has revolutionized security and privacy frameworks in MIoT, ensuring data integrity, trust, and fault tolerance in decentralized systems.

*a) Decentralized Trust Models:* By removing the need for central authorities, blockchain enables trust among heterogeneous IoT devices. Smart contracts facilitate automated operations, reducing human intervention and associated vulnerabilities. Hybrid blockchain systems that combine public and private ledgers are being deployed for scalability and cost-effectiveness [313], [314].

*b) Energy-Efficient Consensus Mechanisms:* Traditional consensus protocols like Proof of Work (PoW) are computationally expensive and unsuitable for energy-constrained MIoT devices. Lightweight mechanisms such as Proof of Stake (PoS), Delegated Proof of Stake (DPoS), and Byzantine Fault Tolerance (BFT) are increasingly adopted for their low energy footprint and faster consensus achievement [315], [316].

*c) Blockchain-Integrated Security Frameworks:* The integration of blockchain with AI enhances MIoT security by enabling predictive threat detection and automated response mechanisms. Blockchain-based identity management systems ensure secure and authenticated device interactions, reducing the risk of impersonation attacks [317], [318].

These advanced tools collectively enable the MIoT ecosystem to meet the demanding requirements of next-generation networks, paving the way for enhanced scalability, energy efficiency, and security. Table 10 provides a comparative analysis of some existing scalable tools for massive IoT.

## VI. EMERGING TRENDS AND FUTURE SOLUTIONS

The rapid expansion of massive IoT presents a myriad of scalability challenges that necessitate innovative approaches. Several emerging trends and future solutions are increasingly gaining traction in the literature, signaling a transformative shift towards more efficient, resilient, and adaptive MIoT systems.

### A. Artificial Intelligence and Machine Learning

The integration of AI and ML techniques is expected to be pivotal in addressing scalability and operational efficiency in MIoT systems. AI and ML enable dynamic resource allocation, predictive maintenance, and efficient data management, which are crucial in dense MIoT environments [119], [319]. For instance, AI-driven predictive analytics can optimize network resources by forecasting traffic patterns, allowing for preemptive resource distribution and reducing latency [320]. Such techniques also improve energy efficiency by identifying low-activity periods and adjusting device operation accordingly [321].

ML algorithms, particularly reinforcement learning and federated learning, have shown promise for real-time network adaptation. These algorithms can respond to fluctuating data demands, facilitating on-the-fly adjustments in MIoT network configurations and enhancing scalability [25], [322]. For instance, reinforcement learning can be utilized to manage network congestion by dynamically rerouting traffic to less

TABLE IX
COMPARATIVE ANALYSIS OF SCALABLE TOOLS IN MIoT.

| Tools | Reference | Network Scalability | Data Management | Energy Efficiency | Security and Privacy | Advantages | Disadvantages |
|---|---|---|---|---|---|---|---|
| Game Theory for Resource Allocation | [266], [267] | ✓ | ✓ | ✓ | | Optimizes resource allocation; adaptive in competitive/cooperative scenarios | Complexity in real-time implementation; computation-intensive |
| Mean Field Game Theory | [271], [285] | ✓ | ✓ | ✓ | ✓ | Handles massive device interactions efficiently; decentralized decision-making | Complexity in mathematical modeling; convergence issues in some scenarios |
| Artificial Intelligence and Machine Learning | [25], [284], [289] | ✓ | ✓ | ✓ | ✓ | Enables intelligent resource allocation, adaptive protocols, and privacy-preserving federated learning | Requires high computational power; challenges in model scalability for constrained devices |
| Queueing & Graph Theory | [118], [293] | ✓ | ✓ | ✓ | | Models large-scale network dynamics; improves throughput and congestion management | Complexity in real-world implementation; potential scalability issues in large-scale scenarios |
| Optimal Transport Theory | [302], [303] | ✓ | ✓ | ✓ | | Efficient data flow optimization; enhances load balancing and federated learning | Computational cost in large-scale networks; requires precise modeling |
| Optimization & Control Theory | [310], [311] | ✓ | ✓ | ✓ | ✓ | Ensures stability in large-scale MIoT; supports adaptive control and optimization | Requires accurate system modeling; high computational complexity in real-time control |
| Blockchain and Distributed Ledger Technology | [314]–[316] | ✓ | ✓ | | ✓ | Decentralized trust; robust against cyberattacks; supports secure data exchange | High latency; energy consumption in some consensus mechanisms |

**Legend**: ✓ indicates the technique's impact in each category.

congested paths [323]. Federated learning, on the other hand, supports privacy-preserving data analysis by processing data locally on devices, which reduces data transfer demands and enhances security [324].

Despite these benefits, challenges remain. AI implementations often require high computational power, which can be difficult to achieve in resource-constrained IoT devices. Additionally, the need for extensive, high-quality training data limits the applicability of some AI models in MIoT environments. Future research aims to address these limitations by developing lightweight AI models suitable for low-power devices, thereby expanding the practical deployment of AI and ML in MIoT [325]. Another promising approach is the hybrid use of cloud and edge AI to distribute computational loads and enhance real-time decision-making across the network [326].

### B. Edge and Fog Computing

The paradigms of edge and fog computing have become critical in managing the scalability of massive IoT systems. These architectures reduce latency and bandwidth demands by processing data closer to the source, minimizing the strain on centralized cloud infrastructures and enabling real-time analytics [327], [328]. This local processing capability is essential for time-sensitive applications, such as autonomous vehicles, smart manufacturing, and remote healthcare monitoring, where delays can impact safety and performance [329].

Recent studies have emphasized the need for optimizing the placement of edge and fog resources to achieve effective load balancing and network resilience [330]. Strategies include using SDN and Network Function Virtualization (NFV) to dynamically allocate resources based on real-time network traffic, ensuring seamless operation under high data loads [331]. Additionally, energy-efficient fog nodes and AI-driven algorithms can predict traffic patterns, further enhancing scalability and energy management in large-scale MIoT environments [332]. Future advancements in edge and fog computing will likely focus on optimizing these deployment strategies and leveraging AI to adapt to network demands in real-time, which is essential for the sustainable growth of MIoT applications.

### C. Network Slicing and 5G/6G Technologies

The deployment of 5G and the anticipated roll-out of 6G technologies present substantial opportunities for enhancing MIoT scalability. Network slicing, which allows multiple virtual networks to operate on a single physical infrastructure, enables tailored connectivity solutions for various MIoT applications [333]. This capability allows IoT devices with diverse requirements—from low-power sensors to high-bandwidth applications—to coexist without interference, effectively optimizing network performance [334].

In addition, 5G and future 6G technologies introduce URLLC and mMTC, addressing scalability by supporting dense device connectivity and high-speed data transfer [335]. These technologies aim to support real-time applications, such as autonomous vehicles, smart healthcare, and industrial automation, by delivering high-speed connectivity and minimizing latency [336].

Moreover, the integration of AI and ML into network infrastructures is expected to be a focus of future research, enabling dynamic network management that adapts to real-time conditions and user demands [337]. Through AI-driven network optimization, MIoT systems could allocate resources proactively, thus improving scalability and user experience across diverse applications. Future advancements in 6G, such as the use of THz frequencies, may further enhance network slicing capabilities, allowing for unprecedented data speeds and supporting MIoT ecosystems [338].

### D. Decentralized Protocols and Blockchain

Decentralized protocols and blockchain technology offer promising solutions to enhance both the security and scalability of MIoT systems. By enabling secure, distributed data exchanges without the need for centralized authorities, blockchain helps eliminate single points of failure, reduce latency, and improve data integrity [339], [340]. This approach is particularly useful in large-scale IoT deployments where traditional centralized systems can become bottlenecks due to high data volumes and network traffic.

The integration of lightweight blockchain solutions is a focus of recent research, as these frameworks aim to accommodate the resource constraints of IoT devices. For instance, new consensus mechanisms, such as PoS and DAG-based blockchains, offer reduced computational requirements compared to PoW, making them more feasible for IoT applications [341], [342]. Furthermore, advancements in interoperable blockchain systems enable communication across different platforms, fostering broader adoption in MIoT ecosystems [343].

However, challenges persist, including the need for improved scalability in blockchain networks and interoperability across various decentralized systems. Future research is anticipated to focus on hybrid blockchain architectures that leverage multiple consensus mechanisms, such as combining PoS with DAG-based approaches, to optimize scalability while ensuring robust security [344], [345]. Such models can enhance the feasibility of large-scale MIoT systems by improving transaction throughput and energy efficiency, addressing key demands in high-density IoT networks.

### E. Sustainable Practices and Energy Harvesting

Sustainability is a central focus in the development of MIoT, with a growing emphasis on energy-efficient designs and renewable energy sources. Energy harvesting technologies such as solar, wind, and kinetic energy are increasingly integrated into MIoT devices, enabling them to operate autonomously and reduce dependence on non-renewable power sources [346], [347]. These technologies allow devices in remote or off-grid locations to remain functional without frequent battery replacements, supporting long-term, sustainable IoT deployment [138].

Recent advancements also include piezoelectric and thermoelectric energy harvesting, where devices generate power from environmental vibrations or temperature differences, making them particularly useful in industrial IoT applications [348], [349]. Future MIoT designs will likely focus on low-power architectures that incorporate these energy-harvesting techniques, ensuring scalability without sacrificing environmental sustainability [350].

Additionally, research suggests that by combining energy harvesting with efficient data transmission protocols and low-power communication standards, MIoT networks can maintain high performance while minimizing environmental impact. This approach will be essential as MIoT systems scale to support millions of devices across various applications, ensuring a balance between technological growth and ecological responsibility [39], [351].

In summary, the landscape of MIoT is rapidly evolving, with various emerging trends and solutions poised to address scalability challenges. The integration of AI, edge computing, network slicing, decentralized protocols, and sustainable practices will be key to enabling efficient, resilient, and scalable MIoT systems in the future.

### F. Convergence of Metaverse and Massive IoT

The convergence of the Metaverse and MIoT is poised to redefine digital interactions by creating immersive, data-rich environments that seamlessly blend the physical and virtual worlds. This integration leverages the extensive connectivity of MIoT devices to enhance the Metaverse's functionality, offering real-time, context-aware experiences across various sectors.

- Enhanced Immersive Experiences: MIoT devices serve as a bridge between the physical and virtual realms, enabling the Metaverse to incorporate real-time data from the physical world. This fusion allows for more dynamic and interactive virtual environments, where users can experience augmented reality (AR) and virtual reality (VR) applications that respond to live physical inputs [107].
- Integration with Advanced Technologies: The synergy between MIoT and the Metaverse is further amplified by the adoption of technologies such as AI, 5G, and blockchain. AI facilitates intelligent data processing and decision-making within the Metaverse, 5G ensures the high-speed connectivity required for seamless interactions, and blockchain provides secure data transactions and asset ownership verification [352].
- Development of Digital Twins: By integrating MIoT data, the Metaverse can host digital twins—virtual replicas of physical entities—that mirror real-time changes and behaviors. This capability is invaluable for simulations, predictive maintenance, and optimizing operations in industries such as manufacturing and urban planning [109].
- Advancements in Edge Computing: To manage the vast amounts of data generated by MIoT devices, edge computing solutions are being developed. Processing data closer to its source reduces latency and enhances the responsiveness of Metaverse applications, leading to more fluid and real-time user experiences [353].

As 6G networks emerge, they are expected to adopt a joint sensing and communication paradigm, further increasing IoT density. This advancement will enable ultra-reliable, low-latency interactions within metaverse environments, enhancing applications such as digital twins and extended reality (XR). However, challenges related to scalability, energy efficiency, and security must be addressed to ensure the successful integration of MIoT within the metaverse [354].

The fusion of the Metaverse and MIoT is an emerging trend that promises to transform how we interact with digital and physical spaces. By embracing this convergence, future solutions will offer more immersive, efficient, and secure applications across various domains, from smart cities to personalized healthcare.

## VII. Concluding Remarks

The MIoT presents an unprecedented opportunity to connect billions of devices across industries, from smart cities to agriculture, offering solutions that enhance productivity, efficiency, and quality of life. However, achieving a scalable MIoT network requires addressing key challenges in network capacity, data management, energy efficiency, and security. This paper has explored a range of current and emerging solutions to tackle these challenges, including LPWAN, 5G, edge computing, and virtualized network infrastructures, alongside emerging approaches like artificial intelligence, blockchain, and machine learning.

As MIoT ecosystems grow, advanced techniques such as grant-free access, NOMA, cooperative communication, and compressive sensing will play a central role in achieving scalability. Integrating these solutions with AI-driven optimization and federated learning offers the potential for a more efficient and adaptive MIoT network. However, future research must focus on refining these technologies for real-world deployment, with particular attention to maintaining energy efficiency, minimizing latency, and strengthening security and privacy.

Ultimately, the path forward for scalable MIoT involves a combination of technical innovation, regulatory support, and cross-industry collaboration. As networks evolve towards 6G and beyond, an emphasis on sustainable practices and adaptable architectures will be essential. The insights provided in this paper can serve as a foundation for future efforts to create robust, secure, and scalable MIoT solutions that meet the demands of an increasingly interconnected world.